\documentclass[sigconf,nonacm]{acmart}

\AtBeginDocument{%
  \providecommand\BibTeX{{%
    \normalfont B\kern-0.5em{\scshape i\kern-0.25em b}\kern-0.8em\TeX}}}

\newcommand{\red}[1]
{{\leavevmode\color{black}#1}}
\newcommand{\blue}[1]
{{\leavevmode\color{black}#1}}

\usepackage{subcaption}

\begin{document}

\title[Navigation Aids in Wide-Area Augmented Reality]{The Impact of Navigation Aids on Search Performance and Object Recall in Wide-Area Augmented Reality}

% \affiliation{%

\author{Radha Kumaran}
\authornote{Both authors contributed equally to this research.}
% \email{rkumaran@ucsb.edu}
% \orcid{0000-0001-7161-3048}
% \email{yujnkm@ucsb.edu}
% \orcid{0000-0003-0903-8999}
\affiliation{%
  % \institution{\nolinebreak[30ex] University of California, Santa Barbara}
  \institution{\normalsize University of California,\linebreak Santa Barbara}
  % \city{Santa Barbara}
  % \state{California}
  \country{\normalsize USA}
  % \postcode{93106}
}

\author{{You-Jin Kim\*}}
\authornotemark[1]  % This reuses the symbol from the first authornote
\orcid{0000-0003-0903-8999}
\affiliation{%
  % \institution{UC Santa Barbara}
  \institution{\normalsize University of California,\linebreak Santa Barbara}
  % \city{Santa Barbara}
  % \state{California}
    \country{\normalsize USA}
  % \country{USA}
  % \postcode{93106}
}

\author{Anne E. Milner}
% \email{amilner@ucsb.edu}
\affiliation{%
  % \institution{UC Santa Barbara}
  \institution{\normalsize University of California,\linebreak Santa Barbara}
  % \city{Santa Barbara}
  % \state{California}
    \country{\normalsize USA}
  % \country{USA}
  % \postcode{93106}
}

\author{Tom Bullock}
% \email{tombullock@ucsb.edu}
\affiliation{%
  % \institution{UC Santa Barbara}
  \institution{\normalsize University of California,\linebreak Santa Barbara}
  % \city{Santa Barbara}
  % \state{California}
    \country{\normalsize USA}
  % \country{USA}
  % \postcode{93106}
}

\author{Barry Giesbrecht}
% \email{giesbrecht@ucsb.edu}
\affiliation{%
  % \institution{UC Santa Barbara}
  \institution{\normalsize University of California,\linebreak Santa Barbara}
  % \city{Santa Barbara}
  % \state{California}
    \country{\normalsize USA}
  % \country{USA}
  % \postcode{93106}
}

\author{Tobias Höllerer}
% \email{thollerer@ucsb.edu}
\affiliation{%
  % \institution{UC Santa Barbara}
  \institution{\normalsize University of California,\linebreak Santa Barbara}
  % \city{Santa Barbara}
  % \state{California}
    \country{\normalsize USA}
  % \country{USA}
  % \postcode{93106}
}
% \author{ANONYMOUS AUTHOR(S) \newline SUBMISSION ID: 4034}
\renewcommand{\shortauthors}{Kumaran and Kim, et al.}

\begin{abstract}

% With head-worn augmented reality (AR) a hotly pursued and increasingly feasible contender paradigm for replacing or complementing smartphones and watches for continual information consumption, we evaluate and compare three different AR navigation aids in a wide-area outdoor user study ($n$=24). 

%Head-worn augmented reality (AR) is a hotly pursued and increasingly feasible contender paradigm for replacing or complementing smartphones and watches for continual information consumption.  Here, we evaluated and compared three different AR navigation aids in a wide-area outdoor user study ($n$=24).An on-screen compass, radar, or in-world vertical arrows respectively alert participants to the location of up to 24 hidden target items. We analyze task performance, participants' movement and head rotation, subjective questionnaire responses and object recall. There were two key findings.  First, all navigational aids enhanced performance compared to the control condition, with some performance benefit and strong user preference for in-world arrows over on-screen compass and then the radar. Second, users recalled significantly fewer physical objects than virtual objects in the environment, suggesting that they may be less aware of their surroundings.  Together, these findings suggest that while navigational aids and AR may enhance search task performance, users may lose situational awareness.

Head-worn augmented reality (AR) is a hotly pursued and increasingly feasible contender paradigm for replacing or complementing smartphones and watches for continual information consumption.  Here, we compare three different AR navigation aids (on-screen compass, on-screen radar and in-world vertical arrows) in a wide-area outdoor user study ($n$=24) where participants search for hidden virtual target items amongst physical and virtual objects.  We analyzed participants’ search task performance, movements, eye-gaze, survey responses and object recall.  There were two key findings. First, all navigational aids enhanced search performance relative to a control condition, with some benefit and strongest user preference for in-world arrows. Second, users recalled fewer physical objects than virtual objects in the environment, suggesting reduced awareness of the physical environment. Together, these findings suggest that while navigational aids presented in AR can enhance search task performance, users may pay less attention to the physical environment, which could have undesirable side-effects.

\smallskip
\textit{ This is a preprint version of this article. The final version of this paper can be found in the Proceedings of ACM CHI 2023. For citation, please refer to the published version.}
\textit{This work was initially made available on the author's personal website [yujnkm.com] in March 2023, and was subsequently uploaded to arXiv for broader accessibility.}

\end{abstract}
% We examined two screen overlay visual cues and a world overlay cue and a baseline without any aids to compare user preference and search behavior while manoeuvring around the physical layout



\begin{teaserfigure}
  \centering\includegraphics[width=0.86\textwidth]{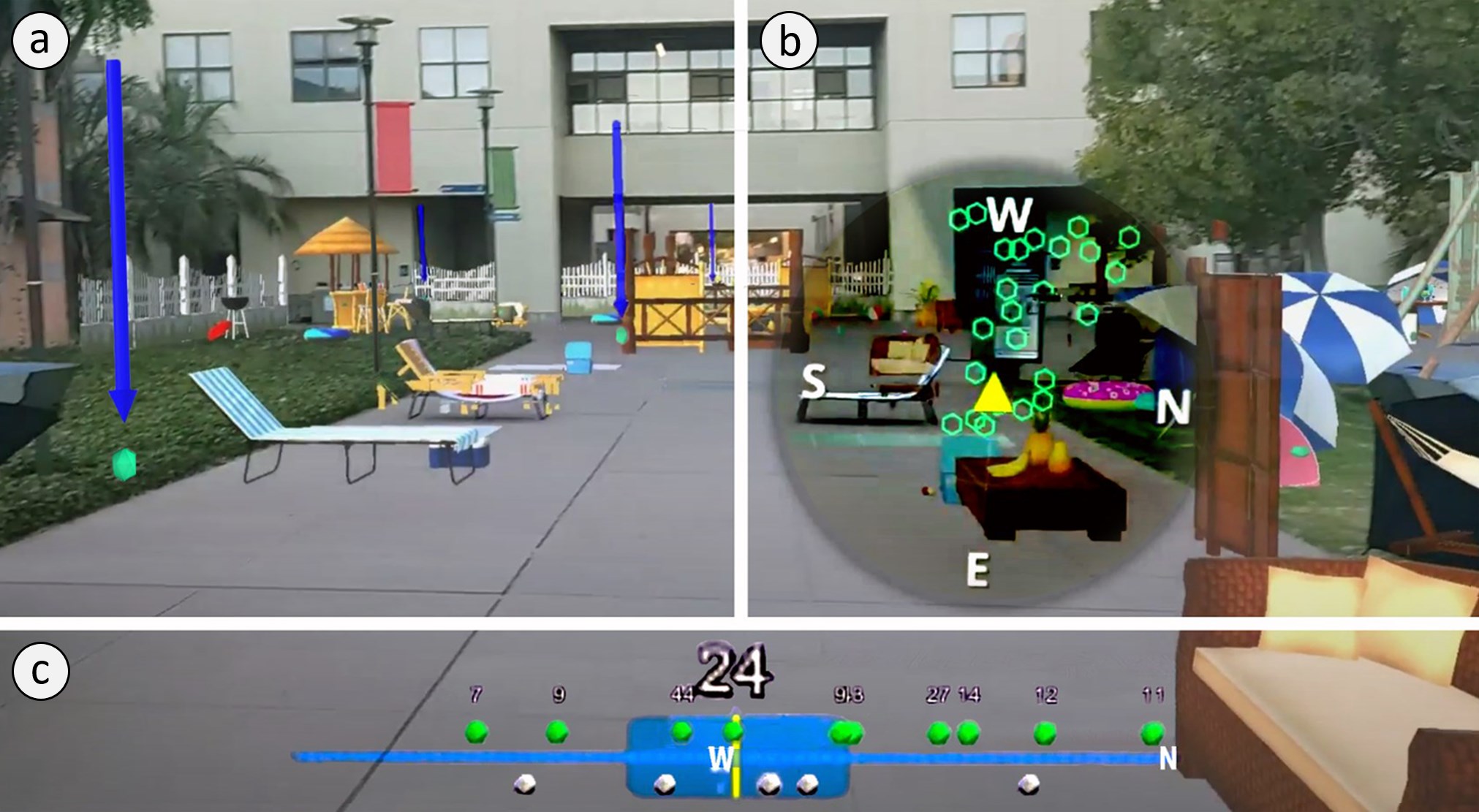} 
  \caption{A view of the experiment environment through the Hololens-2 headset, with examples of each of the three navigation aids (in practice, only one aid is visible at a time). (a) In-world arrows pointing to gems in the environment. (b) On-screen radar indicating positions of gems (green hexagons) relative to the participant (yellow arrow). (c) On-screen compass indicating relative positions of gems positioned in front of the participant (green gems with travel distance in meters displayed above each gem) and behind the participant (white gems). The blue rectangle indicates the forwards field of view and analogous zone behind.}
  \Description{A view of an outdoor university concourse, with virtual and physical objects such as lounge chairs, beach umbrellas, couches and green gems. The screen is divided into three rectangular sections, each representing one of the navigation aids used. The rectangle on the upper left shows long blue arrows pointing vertically downward at the gems. The rectangle on the upper right shows a black disc with annotations representing the user (yellow triangle) and gems (green hexagons). The rectangle across the lower portion of the screen shows a horizontal blue line with green hexagons above the line representing gems in front of the user, and white hexagons representing gems behind the user. A blue rectangle at the centre of the line indicates the field of view and analogous zone behind.}
  \label{fig:teaser}
\end{teaserfigure}

\maketitle

\section{Introduction}

% As basic as search task may be, the wider the search area expands, the user relies on the visual aid such as heads up display radar or compass bar. From the very beginning of AR research, navigational aids displaying relative position, or the distance of the target have been tested to test out the rigidity of the platform. Various heads up display approached has been explored for AR navigational assistant but lacks world base visual assistant that is projected on top of physical layout. We believe world base marking is effective in environment like wide-area outdoor. 

Visual search tasks involve scanning an environment and locating predefined \emph{target} objects in the environment, among other non-target objects (\emph{distractors})~\cite{neisser1963decision}. On a smaller scale, visual search tasks are performed by many people regularly (looking for a specific book on a bookshelf, searching for an ingredient in the pantry) and can usually be completed quickly without any additional assistance. When the search environment is larger however, the complexity of the task increases due to the physical size of the area to be examined as well as the higher number of distracting factors in the environment. In wide-area urban environments, visual search is an important concept that can be supported by mobile augmented reality (AR). For example, a search and rescue team may want to check all entrances and exits of an outdoor mall environment in the aftermath of an earthquake, or a tourist may want to view and be directed to nearby shops or restaurants corresponding to a filtered search query. In such situations, providing assistance in the form of information about the location of the target(s) could help significantly improve the efficiency of the search~\cite{ruddle2009benefits}.

% Navigational aids are often used in video games that require the player to inspect a space and find target(s), both PC-based as well as virtual reality games \red{insert citations}. \red{insert precedent for world-based vs screen-based aids in PC/VR}. 

With augmented reality technology already widely used in mobile devices for navigation (e.g. Google Maps Live View), we see that this technology is not just the future of human-computer interaction - we are already integrating it into existing methods of interaction. However, despite the widespread anticipation of head-worn augmented reality as a potentially primary source of information, there is still limited scientific knowledge on the use and impact of augmented reality outdoors, and in wide areas. Previous work has examined the impact of factors such as lighting on human behavior in outdoors augmented reality \cite{kim2022investigating}, but real-world applications of the technology will involve displaying various types of information to the user in augmented reality to help inform their actions. Head-mounted AR devices are already in use in search and rescue operations~\cite{zhu2021virtual,novetmicrosoft,luksas2022search,julier2000information}, and have also shown potential in applications to education~\cite{wu2013current,garzon2019meta,ibrahim2018arbis}. Although mobile devices are the primary AR interface used in the tourism industry right now~\cite{egger2020augmented}, this could change with the improvement of headset/glasses design and comfort. In these situations, an understanding of how best to present the information is essential to the design of applications that support users both during technology use as well as in the absence of the technology.

The first goal of the experiment reported here was to examine the effect of navigation aids on user performance and behavior in a wide-area augmented reality visual search task. Our analysis offers insights into how users utilize and attend to information presented to them in augmented reality interfaces, and the impacts of different methods of visualizing out-of-view objects. We were especially interested in the difference between navigation aids that presented information in screen space and those that presented information in world space. Our second goal was to assess the difference in user awareness of physical and virtual objects in the environment, since this is an important consideration in the design of augmented reality applications that require the user to interact with both virtual and physical components of their environment. 

In pursuit of the first goal, we collected data from a task that required participants to search for virtual treasure (\emph{gem search}) while also responding to target sounds and (\emph{audio response task}).  To investigate the second goal we conducted an \emph{object recall task} at the conclusion of the study. Participants were required to navigate an outdoor environment augmented with both physical and virtual objects (viewed through the Hololens-2 headset) and search for gems present in the area (\emph{gem search task}). They simultaneously performed a secondary \emph{audio response task}, as a control task to gauge mental load. Three different navigation aids were presented to users as a within-subjects independent variable: an on-screen horizontal compass bar (similar to the Context Compass\cite{suomela2000context}), an on-screen radar, and in-world arrows. The on-screen aids were head-stabilized (attached to the screen, not changing position in the user's field of view, irrespective of head motion), and the in-world arrows were world-stabilized (had a fixed real-world position like any physical object) \cite{billinghurst1999collaborative, hollerer1999situated}. Users were also asked to perform the tasks with no navigation aid, as a control condition. After the experiment, users performed an \emph{object recall task} where they were asked to identify the presence and nature (real, virtual, both real and virtual, or absent) of a list of objects in the environment.

%The current work examined the impact of head-stabilized and world-stabilized navigation aids on user performance in an outdoor wide-area augmented reality search and classification task. 
We expected performance in the search task to be improved when using any of the three navigation aids relative to the no-aid condition, and performance in the audio response task to be impaired with the presence of navigation aids due to an increased focus on the AR information and mechanisms. Among the three navigation aids, we predicted that in-world arrows would result in the greatest benefit.  We did not have any strong predictions regarding any possible difference in benefit between the two on-screen aids. Regarding the recall of physical and virtual objects in the environment, we expected users to recall virtual objects more accurately than physical objects, consistent with previous work \cite{kim2022investigating}.
% To the best of our knowledge, this is the first experiment studying the impact of world-stabilized indicators in wide-area outdoor augmented reality.

The following key insights emerged from the analysis of the data collected in our wide-area outdoor study of navigation aids:
\begin{itemize}
    \item All three navigation aids led to an improvement in performance over the control condition. There were some performance benefits of the arrows over the compass and radar, but no significant differences in performance between the compass and radar.
    \item There was a strong user preference for the arrows over the compass and radar, which is consistent with the performance benefit of the in-world arrows.

\end{itemize}

We also observed potential side effects of augmented reality use: 
\begin{itemize}
    \item Users' significantly lower recall of physical objects in the environment when compared to virtual objects points to a significant shift in attention from the physical world to virtual annotations, which is something that AR application designers need to be aware of. 
    \item Performance in the secondary control task was reduced in the on-screen compass condition, suggesting that the presence of additional on-screen information may impact multitasking ability in augmented reality in certain situations.
\end{itemize}

\begin{figure*}[t]
\centering
  \includegraphics[width=0.7\textwidth]{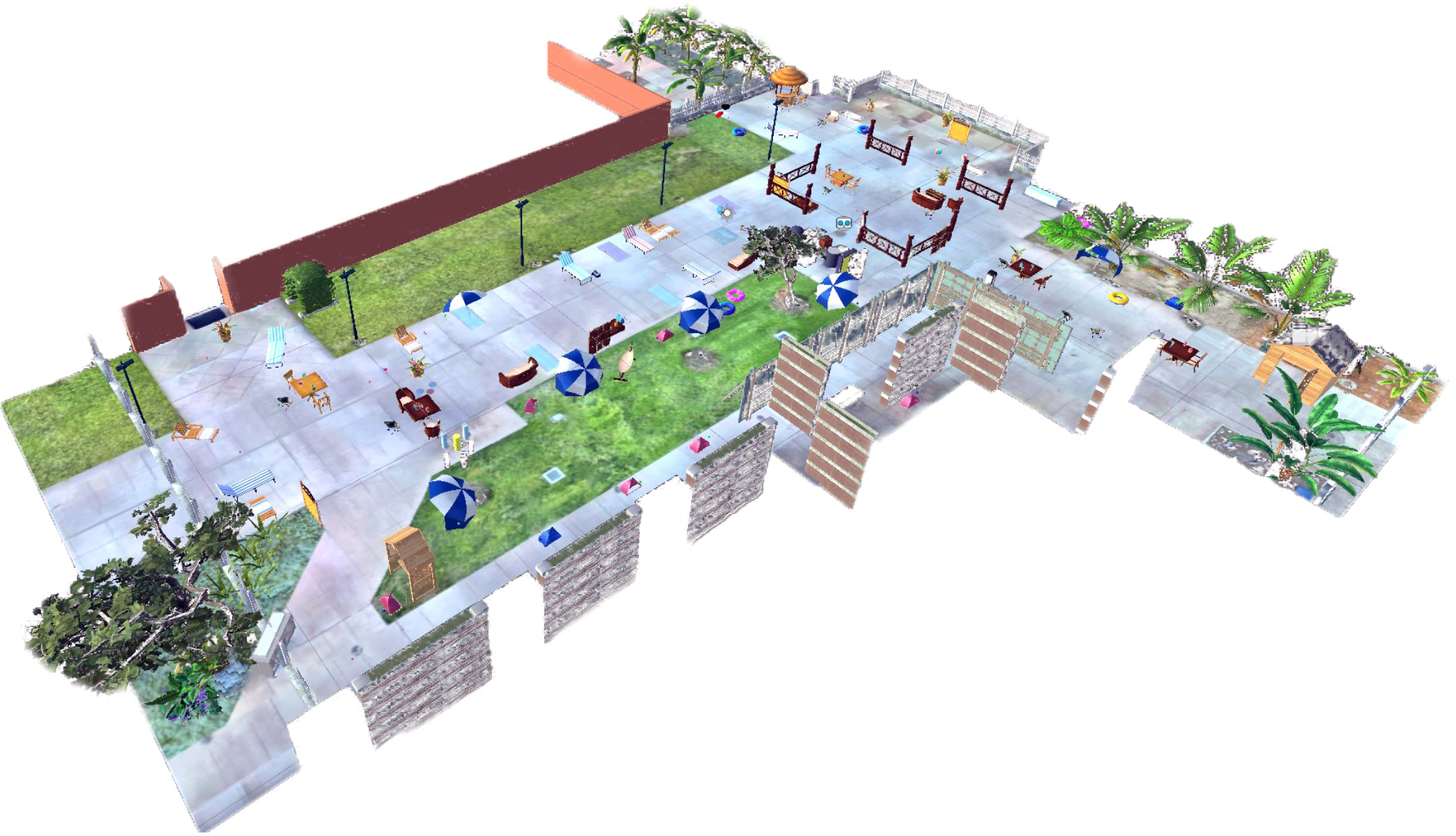}
\caption{A 3D model of the experiment area, augmented with virtual outdoor furniture and other virtual objects.}
\Description{A 3D model of a T-shaped section of a walkway, surrounded by trees and grassy regions. The paved areas of the walkway are augmented with virtual outdoor furniture and other virtual objects.}
\label{fig:model}
\end{figure*}

\section{Related Work}

We discuss previous work in wide-area augmented reality, visual search, navigation aids, and dual-attention, all of which informed the design of the current experiment.

\subsection{Wide-Area Augmented Reality}
Free locomotion and navigation in outdoor wide-area augmented reality has been of interest to the mixed reality domain for a long time~\cite{thomas1998wearable, newman2001augmented, hollerer2004mobile}. However, technical challenges such as sensing, optics, spatial awareness, detection, and recognition technology have led to most user studies so far requiring either controlled environments or external tracking technology. Much of the research in wide-area environments has therefore explored either hand-held mobile augmented reality~\cite{morrison2009like} or virtual reality~\cite{sayyad2020walking}. ARQuake~\cite{thomas2002first} and Human Pacman~\cite{cheok2004human} are augmented reality applications deployed on head-mounted displays that have demonstrated the appeal of experiencing content that is traditionally two-dimensional in augmented reality. It is therefore important to understand how best to display information that helps users perform their tasks in augmented reality, given the strengths and limitations of the technology\cite{kim2022investigating}. This could also prove useful in the design of applications for other purposes, such as collaborative experiences in large-scale environments~\cite{rompapas2019towards, rompapas2018holoroyale}. The present work is one of the first controlled user studies that compares different navigation aids for a search and classification task in outdoors wide-area augmented reality. Further, we examine user perception when interacting with physical and virtual objects.

\subsection{Visual Search}
There is more information in the environment than we can process at any one moment, and attention is the mechanism that allows us to select and prioritize important information that is most relevant to us. Visual search tasks are often used to understand how such information is prioritized when searching for a particular target among distractors in the environment ~\cite {eckstein2011visual}. Typically, the primary metric used to assess the efficiency of search is response time – the time taken to find and correctly respond to the target. Slower response times are indicative of poorer performance and response time has been shown to be affected by the number of distractors and whether targets and distractors share features ~\cite{wolfe1998can}. In the current study, participants searched for known targets among distractor objects in a real-world search environment. The use of navigation aids in visual search can benefit search performance by directing users’ attention to targets in the environment. There have been several previous studies examining the efficiency of different aids that are detailed below.

In augmented reality, guidance to physical or virtual objects outside of the user's current field of view has often been addressed via on-screen arrows \cite{feiner1997touring, thomas1998wearable, gauglitz2014world}. Schinke et al. demonstrated that 3D arrows hinting at off-screen annotations were more effective for memorizing directions to target objects than 2D (top-down) radar maps \cite{schinke2010visualization}. In contrast to these world-stabilized arrows, we opted for vertically aligned 3D arrows (reminiscent of the Hand of God from \cite{stafford2006implementation}) in our study for reasons of unifying and streamlining visual appearance in the presence of many targets (up to 24), for which Schinke's method would have resulted in far too much clutter and confusion. 

If the goal is not just general awareness of peripheral or out-of-sight objects, but the user should also be guided toward a target object, tunnel visualizations have been successfully explored and employed in AR \cite{biocca2006attention,schwerdtfeger2008supporting,shingu2010camera}. For our study, however, we wanted to leave agency of navigational path finding with the participant instead of letting the system decide which next item they should be moving towards. 

\subsection{Navigation Aids}

Various assistants for locomotion navigation and search tasks have been examined in Mixed Reality through handheld devices and head-mounted displays (HMDs)~\cite{mulloni2011user}. A wide variety of attention guiding and navigation aid methods have been explored including visual cues, world markers and and haptic feedback. Many of these studies have been dedicated to people with low vision or limited accessibility~\cite{siu2020virtual, zhao2019seeingvr} to find aids for everyday usability, ease of use and safety. In recent years, several projects have explored visual cues to locate targets outside the HMD’s projected field of view and peripheral vision in navigation search tasks~\cite{biocca2007attention, schinke2010visualization, peck2011evaluation, renner2017attention, gruenefeld2018flyingarrow, chung2021panocue, osmers2020getting}. However, most of these works either studied visual cues in environments with a single search target (or directed users to a specific search target) and hence did not offer much agency in navigational path finding, or compared different visual cues only based on participant-reported measures such as usability, mental workload and preference. We were interested in comparing the impact of each navigational aid on objective search task performance and subjective self-reported measures, as well as examining how each aid interacted with outdoor environmental factors such as lighting.

\paragraph{Radar} Many versions of the radar have been tested in mixed reality, as it is one of the most intuitive methods of spatial navigation assistance and has been widely adopted in first-person shooter games (e.g. Halo, Counter-Strike: Global Offensive). Radar interfaces have been widely explored in many forms including a top-view screen overlay~\cite{chittaro20043d}, an angled radar panel which provides an eagle eye perspective~\cite{gruenefeld2019comparing} and a task specialized radar for messaging on the go~\cite{rantanen2004inforadar}. The performance of two-dimensional radars has been examined in a three-dimensional environment using handheld devices~\cite{burigat2007navigation}. Many different styles of radar interfaces have been tested in virtual and augmented environments~\cite{chittaro20043d, gruenefeld2017eyesee360}. 3D radars have also been used to provide visual guidance in augmented reality headsets which have a limited field-of-view~\cite{bork2018efficient}, and a rotating miniature room layout has been overlaid on peripheral remapping to provide assistance for users with low vision~\cite{zhao2019seeingvr}. Since our experiment focused on providing users with an overview of all the targets in the space, we chose to use a heads-up two-dimensional radar.

\paragraph{Compass} The use of a compass bar as a navigation aid has been explored in games (e.g. The Elder Scrolls V: Skyrim, PlayerUnknown's Battlegrounds) as well as virtual environments in many forms~\cite{osmers2020getting, buchmann2008directional} - as a horizontal compass focused on highlighting targets outside the field of view~\cite{suomela2000context}, top-down and in-world compasses for human-robotic interaction ~\cite{humphrey2008compass}, and with an omnidirectional panoramic view for targets outside the field of view~\cite{chung2021panocue}. Furthermore, variants of the compass bar have been tested in AR indoor environments, such as X-ray vision, 3D compasses and the forward-up compass~\cite{osmers2020getting, rizzi2022design, darken1999map}. In this work we adapted the horizontal Context Compass \cite{suomela2000context} to highlight all targets in the environment, and extended to a 360$^{\circ}$ range based on recommendations in previous work\cite{buchmann2008directional}. 

\paragraph{In-world guides}

In-world navigation guides are situated in the physical layout of the world, and have been found to be more intuitive to use in virtual environments~\cite{wallgrun2020comparison, chung2021panocue}. Many versions of arrows have been used to indicated target locations in virtual environments~\cite{chittaro20043d, burigat2007navigation}, both within the field-of-view\cite{zhao2016cuesee} and outside it~\cite{gruenefeld2018flyingarrow}. SeeingVR's tunnel vision in ``Peripheral Remapping" allows users to visualize their position in relation to the layout of the environment, by providing a miniature version of a wide field navigation guide~\cite{zhao2019seeingvr}. In-world guides have also been shown to be effective at facilitating navigation for users with low vision~\cite{zhao2020effectiveness, zhao2019designing}. While the use of spatial arrows in outdoor augmented environments has been explored~\cite{hollerer2004mobile, gruenefeld2018flyingarrow, zhao2020effectiveness} they have not been well studied in the context of search tasks in outdoor AR.

An analysis of the work in this space informed our choices of navigation aid when designing this study. The use of assistive interfaces to navigate environments for targeted search showed clear benefits over navigation in the absence of these interfaces~\cite{ruddle2009benefits}. In addition, spatial guidance has been preferred over on-screen display interfaces~\cite{wallgrun2020comparison}. We test these ideas in an outdoors wide-area augmented reality environment.

\begin{figure*}[t]
\centering
  \includegraphics[width=0.9\textwidth]{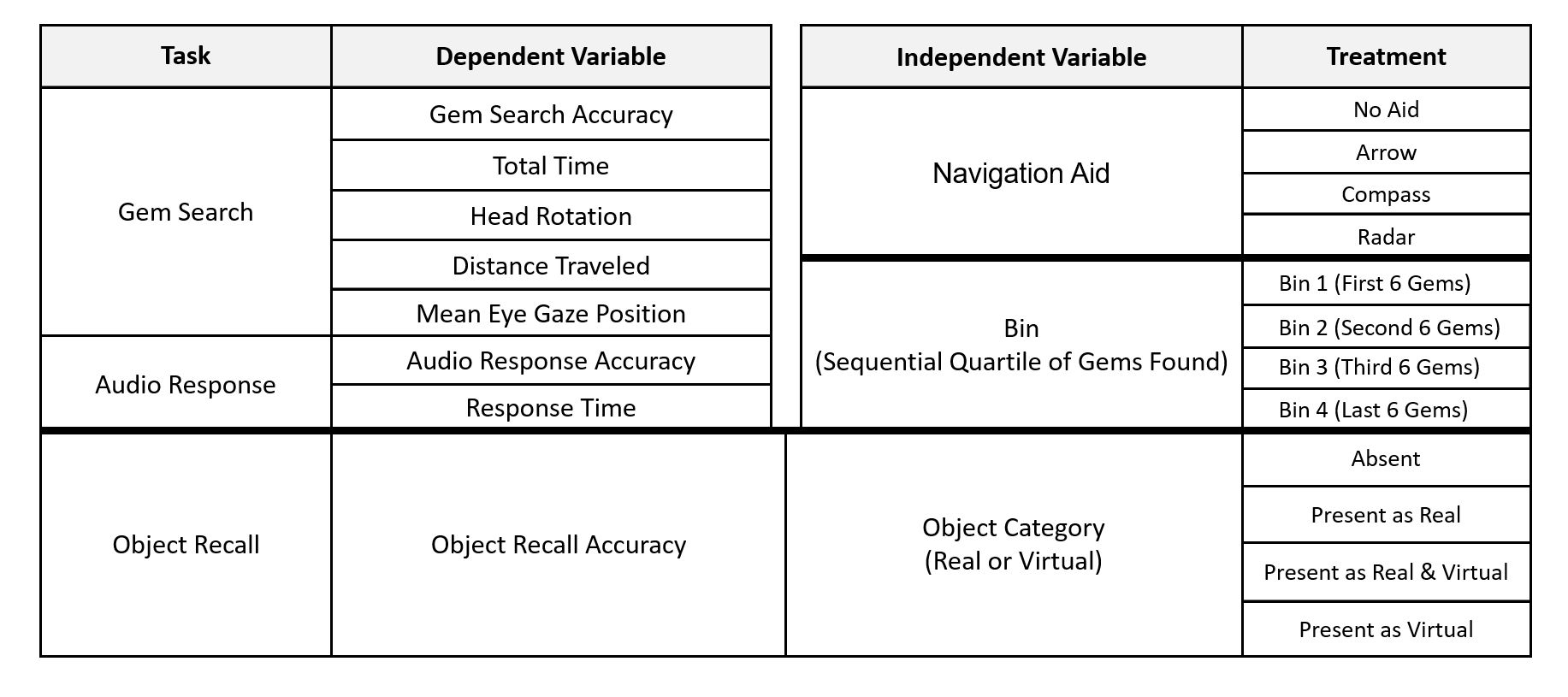}
\caption{A summary of the different design parameters of this experiment.}
\Description{A summary of the different design parameters of this experiment, including tasks, dependent variables, independent variables and treatments for each independent variable. Details are described in the main text. }
\label{fig:design}
\end{figure*}

\subsection{Divided-attention}
Dual-task paradigms involve completing two tasks simultaneously. As human attentional resources are limited, it has been suggested that performing multiple tasks at once can result in a decrease in performance in one or both tasks. This is often observed as an increase in response time and an increase in error rates \cite{pashler1989chronometric}. In the current study, participants completed the experiment under dual-task conditions where participants completed a gem classification task while also responding to an auditory task. We were interested in investigating whether there was an impact on auditory task performance when participants were using different navigation aids to assist with the gem classification task.

\section{Experiment}
In order to study the impact of navigation aids in wide-area augmented reality, we designed a visual search task that required participants to find all gems present in the outdoor experiment area. This task was adapted from a previous experiment reported in the literature~\cite{kim2022investigating}. 

\subsection{Tasks}
Users performed two tasks during the experiment, and one after. The gem search task was designed to encourage free locomotion in the wide-area environment while using navigation aids, and the audio response task was introduced to measure the cognitive impact of using navigation aids. The object recall task was designed to gain insight on the impact of augmented reality on user attention to different parts of their environment (physical or virtual objects). 

\paragraph{Gem search} During the experiment, participants were asked to find all 24 gems in the space, using the navigation aid if one was present. We also introduced a discrimination task, where they were asked to classify each gem into one of four categories, depending on their orientation (vertical, horizontal) and texture (rough, smooth).  This discrimination task was introduced to ensure participants would walk right up to each hidden gem, as the texture and orientation discrimination could not be resolved from further away.

\paragraph{Audio response} This task was introduced as a control task during the experiment, to gauge mental load when performing the gem task with the different navigation aid conditions. Five words were played in a random order for the duration of the participants' search (with a 2-5 second delay between words), and the participant had to respond to each occurrence of their assigned target word.

\paragraph{Object recall} Post-experiment, participants were tested on their recall of the environment by classifying each of a list of objects into one of four categories: absent, present as a real object, present as a virtual object, and present as both a real and virtual object. 

\subsection{Design}

The design of this experiment is summarized in Figure \ref{fig:design}.
In order to analyze participant performance we recorded head position, orientation and eye gaze for the entire duration of each trial, as well as participant responses to both gem and audio task. We also collected object recall responses after the experiment.

\begin{figure*}[t]
\centering
  \includegraphics[width=0.7\textwidth]{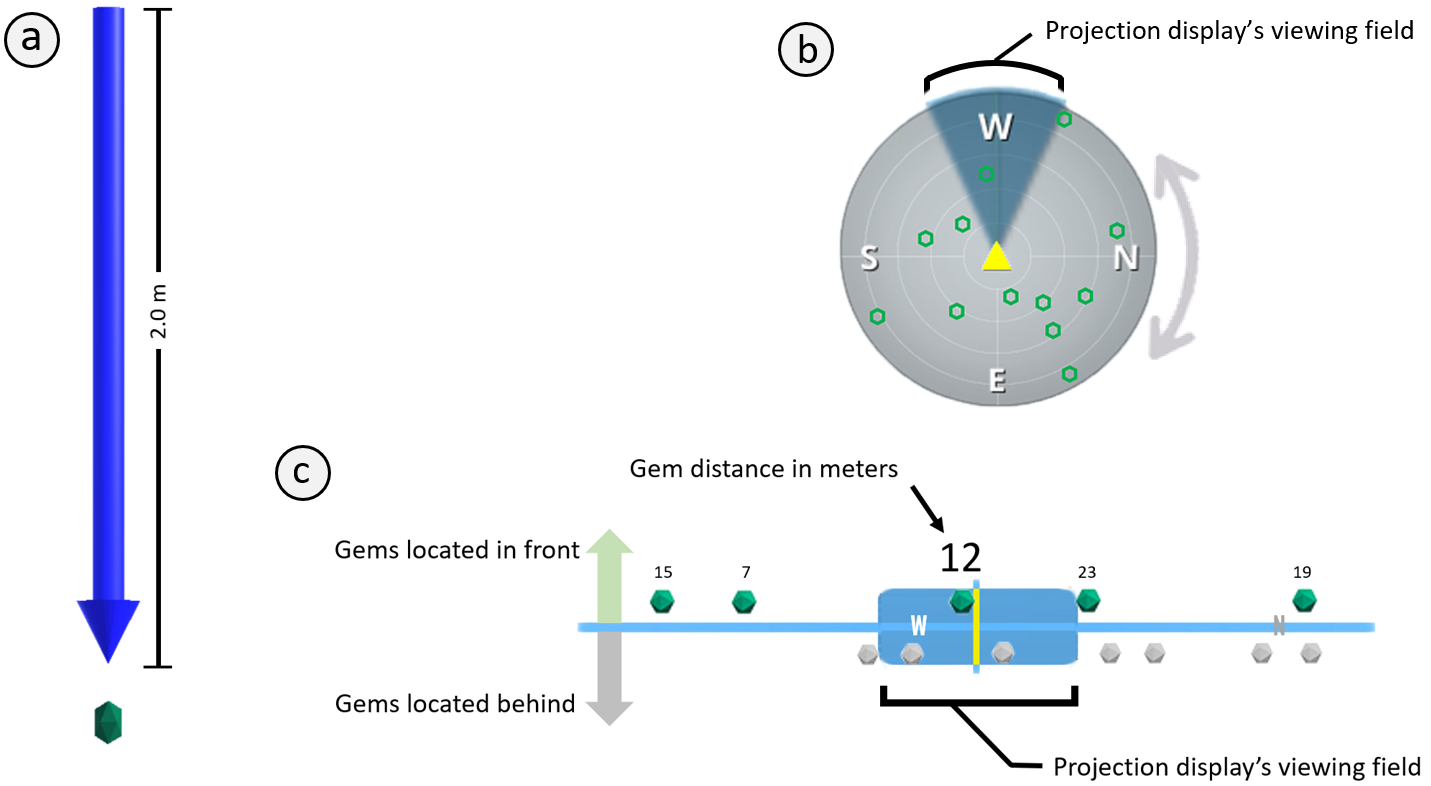}
\caption{Visuals of the three navigation aids the participants saw. Images not to scale. (a) In-world arrows, vertically above each gem. (b) On-screen radar, with a forward-up design and gems indicated by hollow green hexagons. The yellow triangle in the centre indicates the position of the user in the space, and the blue cone represents the field of view. (c) On-screen horizontal compass, green hexagons above the line indicate gems in front of the user (with distance to each gem above the hexagon) and white hexagons below the line show gems behind the user.}
\Description{Visuals of the three navigation aids the participants saw. (a) A 3-dimensional blue arrow, 2 metres long, pointing vertically downward at a green gem. (b) A black disc with a yellow triangle at its center representing the user, and green hexagons representing gems randomly positioned throughout the disc. A blue cone positioned vertically with its narrow end at the center of the disc, and the wider end vertically above the center of the disc, represents the field of view. (c) A horizontal blue line with green hexagons above the line representing gems in front of the user, and white hexagons representing gems behind the user. A blue rectangle at the centre of the line indicates the field of view and analogous zone behind.}
\label{fig:aids}
\end{figure*}

\subsubsection*{Dependent Variables} 

% gem search: 
% - gem search accuracy
% - task time 
% - head orientation 
% - distance walked 
% - eye tracking
% - user preference

% audio task: 
% - accuracy
% - response time 

% object recall task 
Performance in the gem task was initially measured as the fraction of gems correctly discriminated (\emph{gem search accuracy}) out of the 24 targets in each trial. We also measured global behavioral metrics such as the total \emph{time taken} to complete each trial; total \emph{head rotation} which was measured as the accumulated quaternion distance norm; and the total \emph{distance traveled} during each trial. Each of these behavioral metrics were divided into four bins based upon number of gems that had been found so far in the trial, gems found: 1-6, 7-12, 13-18, and 19-24. \emph{Mean eye gaze position} in each trial was also used to estimate user attention to the on-screen navigation aids.

Performance in the audio task was measured as the fraction of target words correctly responded to (\emph{audio response accuracy}). Participants' mean  \emph{response time} in the audio task was also measured, as the time between the onset of the target word and the participants' response.

Performance in the object recall task was measured as the mean accuracy of classification objects in each of the four object categories (\emph{object recall accuracy}), whether the object was present as a virtual object, physical object, both physical and virtual or whether it was absent.

User preference of the different navigation aid conditions was collected after the study, and each aid was scored based on participant rankings (1-4, 1 being most preferred and 4 least preferred). 

\subsubsection*{Independent Variables} 
We originally planned for two independent variables in the design of this experiment, navigation aid and lighting. The navigation aid was manipulated within-subjects, and lighting was a between-subjects variable. Initial analysis regarding lighting yielded only small impact (all effects are listed in Section \ref{sec:analysis} below), and, markedly, no interaction with `navigation aid', the main variable we were interested in. Thus, we removed lighting as an independent variable in our final analysis. In the analysis of search task performance for the four different aid conditions, we discovered varying task performance for `no aid' and `arrows' along the sequence of targets found, averaged over all participants. The last few targets took much longer and required more head movement to find than earlier targets. That led us to consider task performance metrics in four quartiling "bins", corresponding to the following groups of gems found: the first 6, second 6, third 6 and final 6, with the specific gems in each bin differing for each participant and trial. "Bin" became an additional factor in our analysis.

The order of conditions for the lighting and navigation aid variables were counterbalanced between participants. Each participant experienced a different gem layout for each of their 4 experiment trials, chosen from a set of 8 randomized placements generated beforehand. The order of the gem layouts (including backup layouts) was also counterbalanced between participants using the Latin Square method. \red{Although this counterbalancing was not complete since there were only 24 participants, a random subset of the complete counterbalancing was selected and inspected by the experimenters to ensure that there was no bias towards certain placements at specific positions in the sequence of trials.}

\paragraph{Navigation aid} 
Three different navigation aids were compared in this experiment: an on-screen radar, a version of the on-screen context compass\cite{suomela2000context}, and 3D arrows in the environment (see Section \ref{section:aids}). They are hereafter referred to as \emph{radar}, \emph{compass} and \emph{arrows} respectively. The rationale behind the choice of these three aid conditions was to compare navigation aids in world space (arrows) with those in screen space (radar, compass). Among the on-screen aids, the difference between a two-dimensional egocentric aid (radar) and a linear egocentric aid (compass) was of interest as well. Participants completed a trial with no navigation aid as the control condition. 
% The order that each participant encountered the four conditions in was counterbalanced.

% \paragraph{Gem location} The 24 gems in the environment for each experiment trial, while randomly chosen from a set of 48 possible gems, always included 8 gems in each of the following location categories: occluded by a physical object (hereafter referred to as \emph{physically occluded}), occluded by a virtual object (\emph{virtually occluded}), and not occluded by any object (\emph{floating}).
% \red{images here?}

\paragraph{Bin} Each experiment trial was divided into four sections based on the number of gems found in the trial so far, bins 1, 2, 3 and 4 represented the sections of the trial where the first 6, second 6, third 6 and last 6 gems respectively were found. This was introduced as an independent variable to understand the order effects through the trial, because we discovered varying task performance across the sequence of targets with no aid and in-world arrows.

\paragraph{Light} Each participant experienced one of two outdoor lighting conditions: \emph{ambient natural light} (with no direct sunlight), and \emph{night} (with no natural light, and only artificial environment lighting). Studies in the ambient natural light condition were conducted an hour before sunset and studies in the night condition were conducted an hour after sunset.

\subsection{Navigation Aids}
\label{section:aids}

\paragraph{Arrows}
Each gem had a vertical blue arrow (Figure \ref{fig:aids}a) 2m in length pointing to it, with the lower tip of the arrow 40cm above the gem.

\paragraph{Radar}
The radar (Figure \ref{fig:aids}b) was a circular black semi-transparent area that represented the space around the user, and rotated such that the blue sector vertically above the yellow arrow (which represented the user) corresponded to the region of the experiment area within the user's field of view. This is usually referred to as a \emph{forward-up} design. We modified an existing Unity asset (HUD Navigation System \cite{unityradar}) by adding an indicator of the projection display's viewing field. We had to make significant modifications as the asset was not compatible with VR or Mixed Reality, as stated in the asset's documentation. The position of each gem relative to the user was indicated by a hollow green hexagon in the circular area. When the position of a gem was outside the circular area, the hexagon representing it would remain at the edge of the circular area.

\paragraph{Compass} The compass we used (Figure \ref{fig:aids}c) was based on the Context Compass \cite{suomela2000context}, but modified to display all targets (gems) in the space rather than just those within the field of view. An existing Unity asset (Deluxe Compass Bar \cite{unitycompass}) was adapted to show gems located in the front and rear presenting full 360$^{\circ}$ simulated field-of-view. This ensured a fair comparison with the other on-screen aid (radar), with both aids providing the user with the same amount of information. The blue rectangle indicated the headset's field of view. All gems in front of the user (i.e. within a 180$^{\circ}$ simulated field-of-view) were represented by green hexagons above the blue line, and all gems behind the user were represented by white hexagons below the blue line. Each gem in front of the user also had a label showing the distance in metres to reach it from the user's current position.

\subsection{Apparatus}
The study was conducted outdoors on a university concourse 1456 sq.m. (15,672 sq.ft.) in size, with 185 sq.m. (1991 sq.ft.) of lawn area, which participants were discouraged (but not forbidden) from walking on. Participants viewed the augmented environment using Microsoft's HoloLens-2 headset. The environment was modeled and augmented with virtual and real objects such as lounge chairs, shade tents, and beach umbrellas in Unity (see Figure \ref{fig:model}). Participant responses were recorded using a bluetooth gamepad controller, and the experimenters controlled the study using a bluetooth keyboard. For studies conducted after dusk, the area was illuminated with 14 ring lights strategically placed to ensure sufficient illumination throughout the experiment area. Experimenters monitored the participants' progress using a custom web app, accessed through a mobile device.

\subsection{Participants}
Participants were 24 adult volunteers (11 female, 20 right-handed) between the ages of 18 and 31 ($M$=23.16, $SD$=3.42). They were compensated at a rate of \$15 per hour. All participants reported normal or corrected-to-normal vision, with 7 participants using vision correction. 14 participants had never used AR before, and only 3 had used it more than 10 times.

\subsection{Procedure}
Participants were first asked to fill out a demographic questionnaire. They then calibrated the study headset for their eye gaze, and proceeded to complete a short training module (on a different headset) introducing them to the experiment while the experimenter set up and aligned the study environment. In the training module (which was completed in an area outside the study space, also augmented with virtual objects), participants were introduced to the two tasks and three navigation aids. They practiced the two tasks in all four experiment conditions exactly as they would in the main study, but only classified a couple of gems in each condition. Once the training was complete, the participant put on the study headset and completed four sets of trials (one for each navigation aid condition, and one for the control condition). Each set comprised a practice trial (where the participant had to find any four gems out of 12 present in the environment), and an experiment trial (where they needed to find all 24 gems, or as many as possible within the time limit of 10 minutes). Participants were informed that the practice trial was for them to get used to the navigation aid, to reduce the impact of the learning effect on their performance. Participants were given a 10-second break between trials. In case of tracking loss, which occurred in 12 of the 96 (24 x 4) trials, the trial was abandoned and a backup trial (with a different gem layout for the same navigation aid condition) was run instead. Once the participant completed all 4 sets of trials they answered a post-study questionnaire that collected feedback on the ergonomics and enjoyment of the experience, as well as tested their recall of objects present in the environment and ranked their preference for each of the aids. 

\subsection{Analysis}
\label{sec:analysis} 

All the data was processed in a custom playback software that allowed replay of each participants' trajectory and actions in a virtual model of the study environment. This playback was monitored to ensure that there were no inconsistencies in the collected data, and also to check for any segments of the trial where registration inaccuracies occurred. Such segments were observed in 5 of the 96 trials, and excluded from the analysis. 

 To assess the impact of each navigation aid (no aid, arrows, radar, and compass) and bin (1, 2, 3, 4) on our global behavioral metrics we conducted repeated measures ANOVAs. To assess the impact of navigation aid on accuracy and response time for the audio task one-way repeated measures ANOVAs were used. A similar one-way repeated measures ANOVA was used to examine the effect of navigation aid on mean classification accuracy in the object recall task. All ANOVAs were conducted using the {\small rstatix} package in R. The assumption of homogeneity of variance for ANOVA was not violated and the Greenhouse-Geisser correction was used if the assumption of sphericity had been violated. Pairwise comparisons using the Bonferroni correction were used to follow-up significant main effects and interactions. 
 
 To analyze ranked preference scores for each navigation aid Friedman's test was used, which is a non-parametric equivalent of repeated measures ANOVA appropriate for the analysis of ranked scores. Pairwise comparisons were conducted using the Wilcoxon signed rank test. 

%To assess whether the two lighting conditions had an impact on our results, all analyses were conducted with lighting as a factor. The only dependent variable found to be affected by lighting condition was distance traveled. A repeated measures ANOVA with the factors: lighting(natural, night), and navigation aid, bin(1, 2, 3, 4) revealed an effect of lighting such that participants traveled further in the night condition than the day condition, $F$(1,22) = 8.79, $p$ < 0.001, $\eta_{}^{\mathrm{2}}$ = .29, $large$. There was also an interaction between lighting and bin, $F$(1.28,28.11) = 6.10, $p$ < 0.001, $\eta_{}^{\mathrm{2}}$ = .22, $large$, such that participants traveled a longer distance in the night compared to the natural lighting condition only in the first and the final bin; bin 1 [$t$(21.6) = -2.51, $p$ < 0.05, $d$ = -1.02, $large$]; bin 4 [$t$(22) = -2.88, $p$ < 0.01, $d$ = -1.18, $large$]. These results suggest that when participants were both adjusting to the task in the initial bin and searching for the final gems they had greater difficulty finding gems in the night relative to the natural light condition. Although this was an interesting finding, it does not relate to our hypotheses, as light was not found to interact with any of the navigation aid conditions across our dependent measures. Therefore, we collapsed over lighting condition and was not included as a factor in our analyses. 

To assess whether the two lighting conditions had an impact on our results, an initial set of analyses was conducted with lighting as a factor. The only dependent variable found to be affected by lighting condition was distance traveled. Results indicated that participants traveled further in the night condition than the natural light condition, but only in the initial and final bin. These results suggest that when participants were both adjusting to the task in the initial bin and searching for the final gems they had greater difficulty finding gems in the night relative to the natural light condition. Although this was an interesting finding, it does not relate to our hypotheses, as light was not found to interact with any of the navigation aid conditions across our dependent measures. Therefore, lighting was not included as a factor in our main analysis reported here. \red {We however report the results with lighting considered as a factor in Appendix \ref{sec:appendix-light}.}

\section{Results}

The results are presented in four sections corresponding to: gem task, audio task, user preferences, and object recall.

\subsection{Gem Search}

Participants were tasked to find all 24 gems placed in the environment, and there was a time limit of 10 minutes per trial. Participants did not find all 24 gems in 20 of the 96 trials(14 no aid; 3 radar; 2 arrows; 1 compass), the minimum number of gems found was 20. Performance was characterized using three behavioral metrics: time taken to find the gems, \blue {head rotation, and walking speed. Additionally, to examine the interaction between time and the other two metrics, we analysed both head rotation and distance traveled accumulated over the entire trial.} Gem discrimination accuracy was also recorded, which, as expected, was subject to a ceiling effect ($Mean$ = 98.50, $SEM$ = 0.006) with participants making few mistakes regardless of aid ($F$(3,69) = 1.31, $p$ > 0.05, $\eta_{}^{\mathrm{2}}$ = .054), gem type ($F$(2,46) = 0.66, $p$ > 0.05, $\eta_{}^{\mathrm{2}}$ = .028), or their interaction ($F$(3.82,87.92) = 0.78, $p$ > 0.05, $\eta_{}^{\mathrm{2}}$ = .033). We purposefully included the discrimination task, however, to ensure participants would walk right up to each hidden gem, as the texture discrimination (and to some degree the gem orientation) could only be resolved from nearby.

%Due to a ceiling effect in search accuracy in the gem search task, we chose to analyse performance 
\paragraph{Time Taken}

\begin{figure}[t]
\centering \includegraphics[width = \columnwidth]{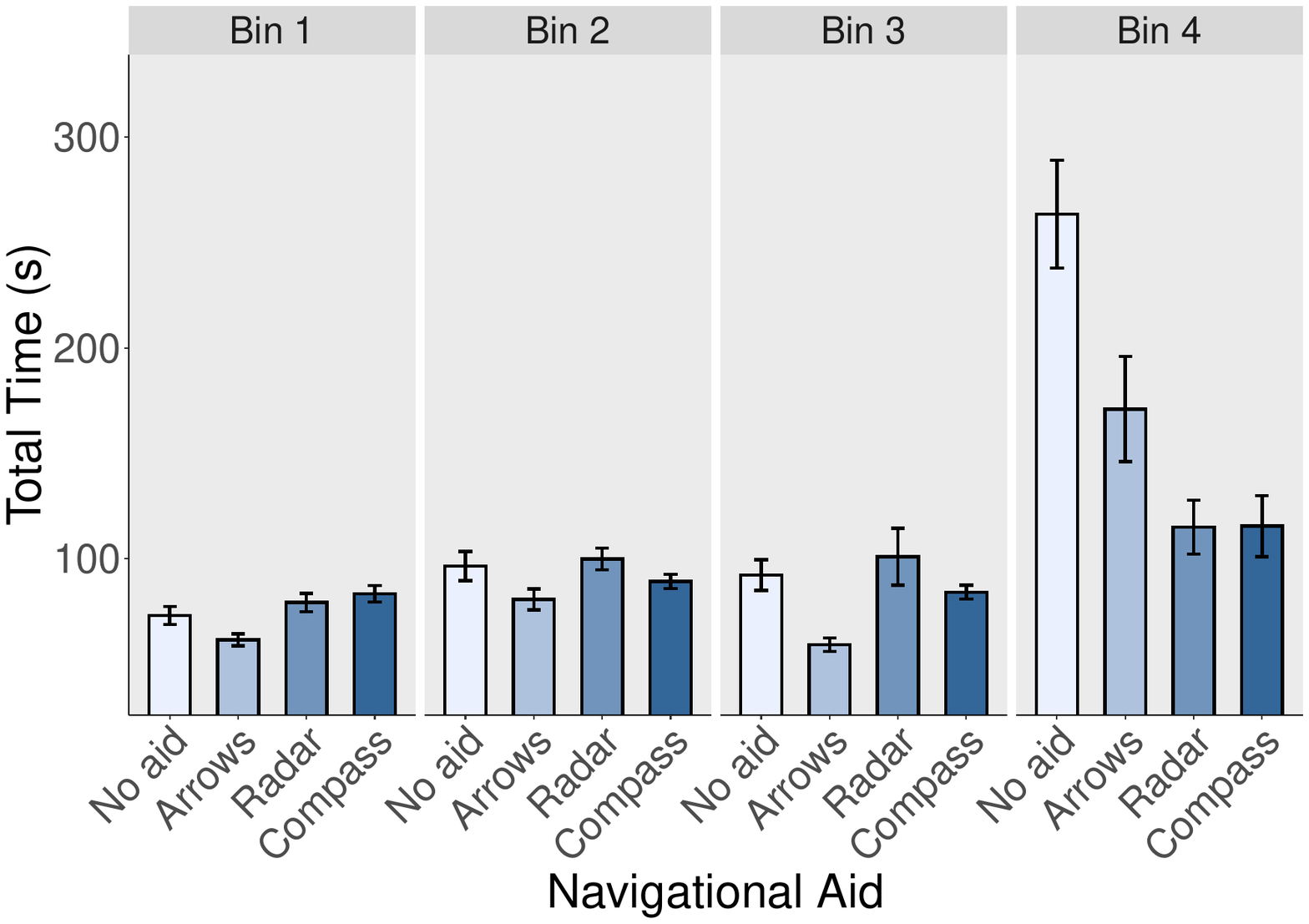}
\caption{Total session time in seconds plotted as a function of navigation aid and bin. Participants took significantly longer to complete a trial with no aid compared to the three aid conditions. Among the three aids, there was a benefit of arrows over the other two aids in the early bins (bin 1 \& bin 3), but this did not extend to the last bin of the trial. Participants also took significantly longer in later bins, as fewer gems remained to be found (bin 4 $>$ bins 2,3 $>$ bin 1), especially in the arrows and no aid conditions. For this and all following figures: Error bars $=$ SEM.}
\Description{Total session time in seconds plotted as a function of navigation aid and bin. Participants took significantly longer to complete a trial with no aid compared to the three aid conditions. Among the three aids, there was a benefit of arrows over the other two aids in the early bins (bin 1 and bin 3), but this did not extend to the last bin of the trial. Participants also took significantly longer in later bins, as fewer gems remained to be found, especially in the arrows and no aid conditions.}
\label{fig:totalTime}
\end{figure}

Figure \ref{fig:totalTime} plots the total average session time per trial as a function of navigation aid and bin (quartiles of gems found: first 6, second 6, third 6 and final 6, with the gems in each bin differing for each participant and trial). Overall, performance was altered by the navigation aid condition ($F$(3,69) = 12.71, $p$ < 0.001, $\eta_{}^{\mathrm{2}}$ = .36, $large$). Participants took significantly longer to complete a trial in the no aid condition compared to the other three aid conditions; no aid versus arrows [$t$(23) = 4.34, $p$ < 0.001, $d$ = 1.16, $large$]; no aid versus radar [$t$(23) = 4.23, $p$ < 0.001, $d$ = 1.12, $large$]; no aid versus compass [$t$(23) = 5.36, $p$ < 0.001, $d$ = 1.46, $large$]. Similarly, overall performance changed as a function of bin ($F$(1.26, 29.66) = 51.62, $p$ < 0.001, $\eta_{}^{\mathrm{2}}$ = .69, $large$), such that participants took significantly longer as they found more gems e.g., bin 1 vs. bin 4 [$t$(23) = -8.68, $p$ < 0.001, $d$ = -2.23, $large$].

In addition to these overall effects of aid and bin, visual inspection of Figure \ref{fig:totalTime} suggests that the effect of aid changed as a function of bin. Among the three aids, participants were faster with the arrows when compared to the screen stabilized aids (compass and radar) in the initial portion of the trial (bins 1 and 3). However, this benefit did not extend to the last section of the trial. They also took longer successively between bin 1, 2, and 4 with arrows and no aid, while there was no such increase in task time with compass and radar. Correspondingly, the interaction between aid and bin was significant ($F$(1.29,29.66) = 51.62, $p$ < 0.001, $\eta_{}^{\mathrm{2}}$ = .33, $large$). In order to break down the interaction we examined the effect of bin at each level of the aid factor. There was an effect of bin in the no aid condition and arrow condition, and pairwise comparisons showed that participants were significantly faster to find gems in the earlier compared to later bins e.g., bin 1 vs. bin 4 ($p's$ < 0.001). There was no main effect of bin in the radar condition. There was a main effect in the compass condition ($F$(1.22,28.04) = 4.09, $p$ < 0.05, $\eta_{}^{\mathrm{2}}$ = .15, $large$), but no comparisons survived correction. In order to further break down this interaction, we conducted a separate ANOVA for the effect of aid at each level of bin. There was a main effect of aid in the first bin ($F$(2.41,55.34) = 6.36, $p$ < 0.001, $\eta_{}^{\mathrm{2}}$ = .22, $large$), which showed that participants took significantly less time to complete the trial in the arrow compared to the radar and the compass condition; arrow versus radar [$t$(23) = -3.76, $p$ < 0.001, $d$ = -1.00, $large$]; arrow versus compass [$t$(23) = -5.34, $p$ < 0.001, $d$ = -1.30, $large$]. There was an effect of aid in bin 2 ($F$(3,69) = 2.86, $p$ < 0.05, $\eta_{}^{\mathrm{2}}$ = .11, $large$), but no comparisons survived correction. There was an effect of aid in bin 3, $F$(1.65,37.93) = 5.05, $p$ < 0.05, $\eta_{}^{\mathrm{2}}$ = .18, $large$). Pairwise comparisons revealed that participants were significantly faster to complete the trial in the arrow condition compared to the three other aids; no aid versus arrows [$t$(23) = 3.97, $p$ < 0.01, $d$ = 1.20, $large$], no aid versus radar [$t$(23) = 3.08, $p$ < 0.05, $d$ = 0.87, $large$], no aid versus compass [$t$(23) = 6.06, $p$ < 0.001, $d$ = 1.55, $large$]. There was also a main effect of aid in bin 4 ($F$(2.18,50.21) = 13.33, $p$ < 0.001, $\eta_{}^{\mathrm{2}}$ = .37, $large$), that showed that participants were significantly slower in the no-aid condition compared to the radar and the compass condition; no aid versus radar [$t$(23) = 5.26, $p$ < 0.001, $d$ = 1.50, $large$]; no aid versus compass [$t$(23) = 5.31, $p$ < 0.001, $d$ = 1.45, $large$]. Participants were no longer faster in the arrows condition relative to the no-aid condition when searching for the final six gems. 

\blue{

\paragraph{Head Rotation}

\begin{figure}[t]
\centering \includegraphics[width = \columnwidth]{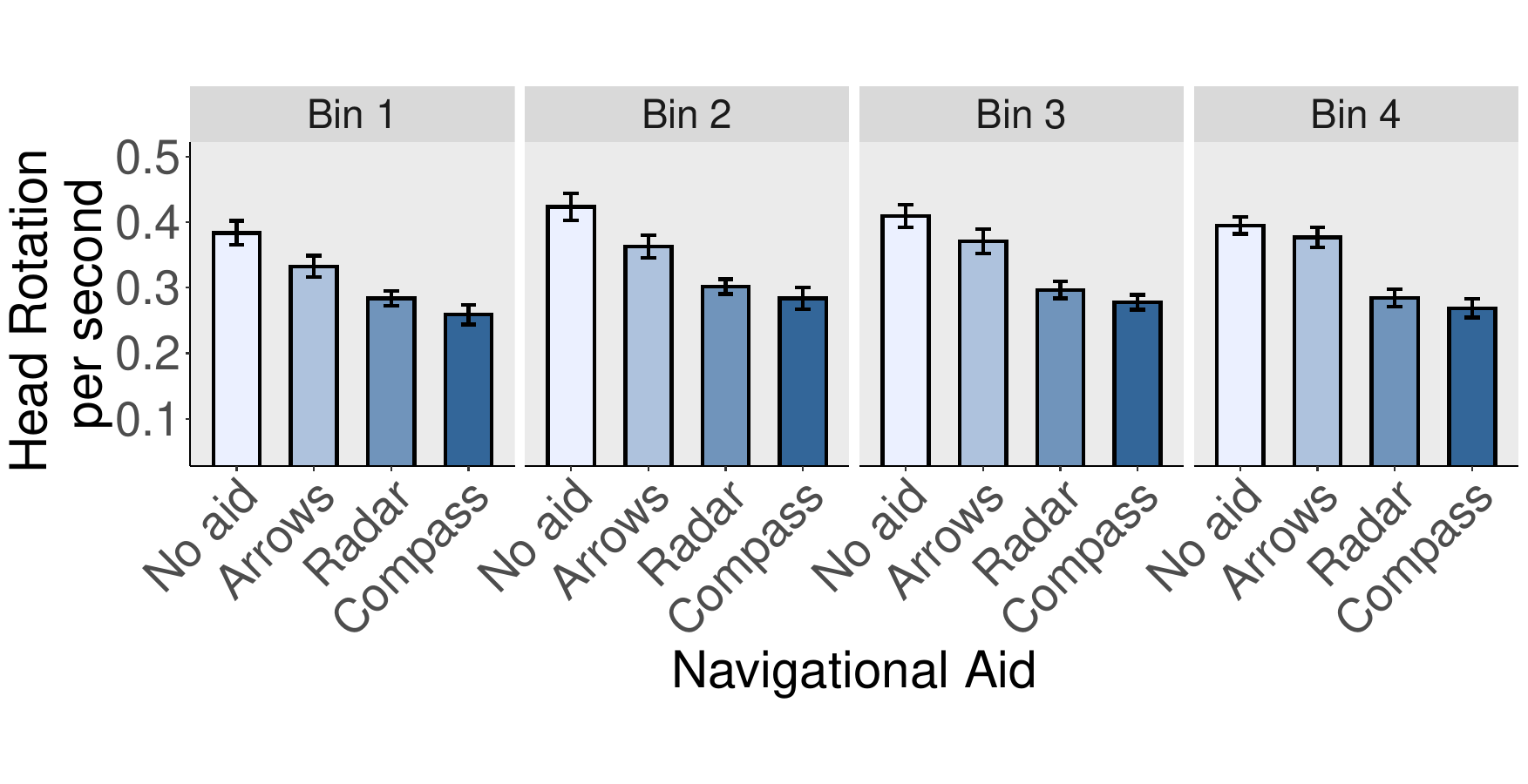}
\caption{\blue{Head rotation plotted as a function of navigation aid and bin. Head rotation was greater in the no aid condition compared to any of the three navigation aids, suggesting that search was less efficient when there was no aid. Head rotation was also greater in the arrows condition compared to compass and radar, suggesting that the on-screen aids enabled a more focused search pattern than the in-world arrows. }}
\Description{Head rotation plotted as a function of navigation aid and bin. Head rotation was greater in the no aid condition compared to any of the three navigation aids, suggesting that search was less efficient when there was no aid. Head rotation was also greater in the arrows condition compared to compass and radar, suggesting that the on-screen aids enabled a more focused search pattern than the in-world arrows.}
\label{fig:unitrotation}
\end{figure}

Figure \ref{fig:unitrotation} shows the head rotation (measured as the norm of the difference of quaternions) per second as a function of navigation aid and bin. Head rotation was greater in the no aid condition compared to any of the three navigation aids, suggesting that search was less efficient when there was no aid. Consistent with this pattern, the ANOVA revealed an effect of navigation aid, $F$(3, 69) = 59.02, $p$ < 0.001, $\eta_{}^{\mathrm{2}}$ = .72, $large$. Head rotation was greater in the no aid condition compared to the other three conditions, suggesting that all three aids benefited search performance by making it easier to discover the next target; no aid versus arrows, [$t$(23) = 4.57, $p$ < 0.001, $d$ = 0.55, $large$]; no aid versus compass, [$t$(23) = 10.50, $p$ < 0.001, $d$ = 1.88, $large$]; no aid versus radar, [$t$(23) = 9.39, $p$ < 0.001, $d$ = 1.70, $large$]. Head rotation was also greater in the arrows aid condition compared to the two on-screen aid conditions; arrows versus compass, [$t$(23) = 7.28, $p$ < 0.001, $d$ = 1.25, $large$]; arrows versus radar, [$t$(23) = 6.52, $p$ < 0.001, $d$ = 1.04, $large$]. This suggests that the on-screen aids enabled a more focused search pattern, and reduced the amount of information users needed to gather from the environment, than the in-world arrows.

Head rotation also increased over time irrespective of the aid condition, which suggests that as participants found more gems over time, search became more difficult. Correspondingly, there was an overall effect of bin, $F$(3, 69) = 9.80, $p$ < 0.001, $\eta_{}^{\mathrm{2}}$ = .30, $large$, such that head rotation significantly increased after bin 1; bin 1 versus bin 2, [$t$(23) = -6.11, $p$ < 0.001, $d$ = -0.44, $large$]; bin 1 versus bin 3, [$t$(23) = -4.47, $p$ = 0.001, $d$ = -0.39, $large$]. 

% We report an analysis of the accumulated head rotation in each bin in Appendix \ref{sec:appendix-behavioral}.
% Pairwise comparisons revealed the following significant differences; no aid versus arrows, [$t$(23) = 5.37, $p$ < 0.001, $d$ = 1.70, $large$]; no aid versus radar, [$t$(23) = 8.10, $p$ < 0.001, $d$ = 2.12, $large$]; no aid versus compass, [$t$(23) = 9.18, $p$ < 0.001, $d$ = 2.91, $large$]. Head rotation also increased over time irrespective of the aid condition, which suggests that as participants found more gems over time, search became more difficult. Correspondingly, there was an overall effect of bin, $F$(1.27, 29.27) = 36.81, $p$ < 0.001, $\eta_{}^{\mathrm{2}}$ = .68, $large$, such that head rotation significantly increased over the course of each bin e.g. bin 1 versus bin 4, $t$(23) = -8.90, $p$ < 0.001, $d$ = -2.54, $large$. 
}

\blue{
\paragraph{Walking Speed}
The participants' walking speed during the task is plotted as a function of navigation aid and bin in Figure \ref{fig:speed}. Overall, results revealed that the speed was affected by aid ($F$(3,69) = 12.57, $p$ < 0.001, $\eta_{}^{\mathrm{2}}$ = .35, $large$). Participants walked significantly faster in the arrows condition compared to the other three aid conditions, suggesting that the arrows benefited search performance; arrows versus no aid, [$t$(23) = 2.94, $p$ < 0.05, $d$ = 0.41, $large$]; arrows versus radar, [$t$(23) = 5.63, $p$ < 0.001, $d$ = 1.11, $large$]; arrows versus compass conditions, [$t$(23) = 3.32, $p$ < 0.05, $d$ = 0.63, $large$]. The increased walking speed in the arrows aid condition compared to the two head-stabilized aid conditions could be an effect of the arrows being in 3D space, thus eliminating the need to spatialize 2D information (from the head-stabilized aids) in 3D space. The increased speed compared to the no aid condition could be because the arrows were easier to spot in the environment compared to the actual targets. Further, since participants needed to spend more time looking at the on-screen aids (rather than the physical environment) in the compass and radar aid conditions, the reduced attention to their 3D environment likely also decreased their walking speed.

\begin{figure}[t]
\centering \includegraphics[width = \columnwidth]{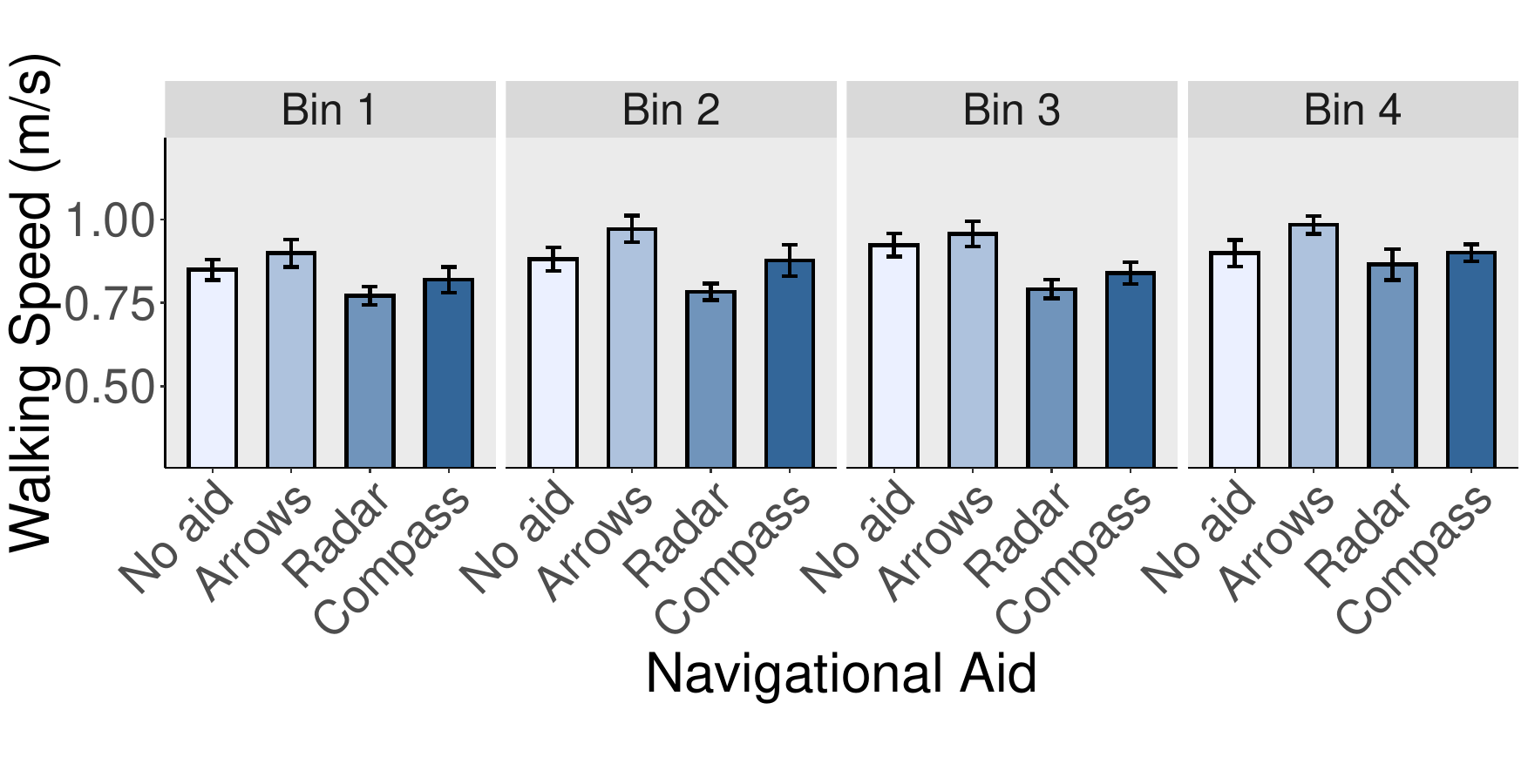}
\caption{\blue{Walking speed in meters per second plotted as a function of navigation aid and bin. Participants walked significantly faster in the arrows aid condition compared to the other three conditions.}}
\Description{Walking speed in meters per second plotted as a function of navigation aid and bin. Participants walked significantly faster in the arrows aid condition compared to the other three conditions.}
\label{fig:speed}
\end{figure}

There was also an overall effect of bin on speed ($F$(2.1, 48.39) = 6.51, $p$ < 0.01, $\eta_{}^{\mathrm{2}}$ = .22, $large$) such that participants walked significantly faster in the later bins compared to the first bin e.g., bin 1 vs. bin 4, ($t$(23) = -3.4, $p$ < 0.05, $d$ = -0.56, $large$). This could be an effect of increased familiarity with the environment in the later bins when compared to the first bin. 

% We report an analysis of the total distance traveled by the participants in each bin in Appendix \ref{sec:appendix-behavioral}.
}

\paragraph{Accumulated Head Rotation}

\begin{figure}[t]
\centering \includegraphics[width = \columnwidth]{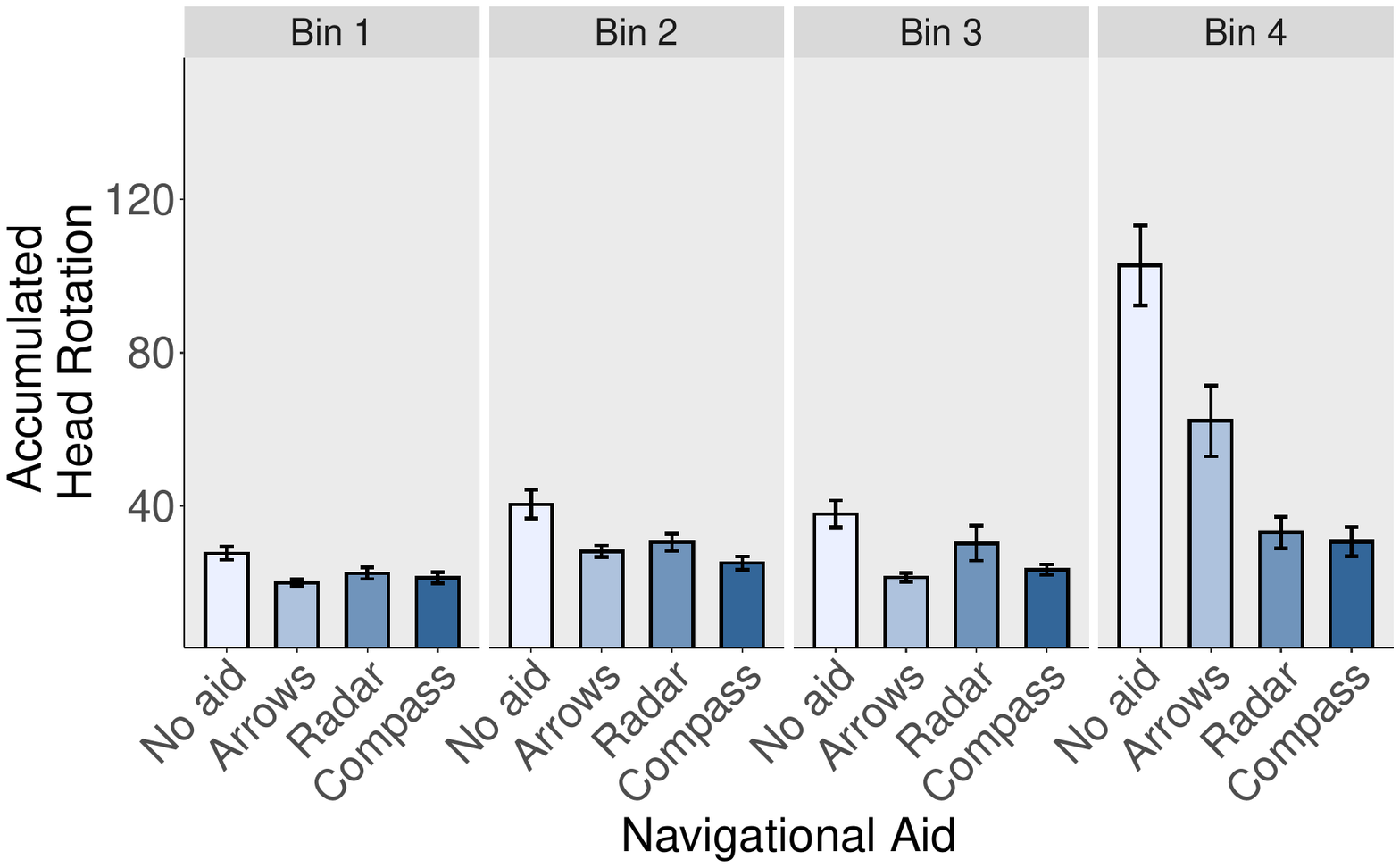}
\caption{Accumulated head rotation plotted as a function of navigation aid and bin. Accumulated head rotation was greater in the no aid condition compared to any of the three navigation aids, suggesting that search was less efficient when there was no aid. The arrows provided a benefit over no aid in bins 1-3, but this did not continue in the final bin.}
\Description{Accumulated head rotation plotted as a function of navigation aid and bin. Accumulated head rotation was greater in the no aid condition compared to any of the three navigation aids, suggesting that search was less efficient when there was no aid. The arrows provided a benefit over no aid in bins 1-3, but this did not continue in the final bin.}
\label{fig:rotation}
\end{figure}

Figure \ref{fig:rotation} shows the accumulated head rotation (measured as the accumulated norm of the difference of quaternions between subsequent captured frames) as a function of navigation aid and bin. Accumulated head rotation was greater in the no aid condition compared to any of the three navigation aids, suggesting that search was less efficient when there was no aid. Consistent with this pattern, the ANOVA revealed an effect of navigation aid, $F$(2.29,52.76) = 36.81, $p$ < 0.001, $\eta_{}^{\mathrm{2}}$ = .60, $large$. Pairwise comparisons revealed the following significant differences; no aid versus arrows, [$t$(23) = 5.37, $p$ < 0.001, $d$ = 1.70, $large$]; no aid versus radar, [$t$(23) = 8.10, $p$ < 0.001, $d$ = 2.12, $large$]; no aid versus compass, [$t$(23) = 9.18, $p$ < 0.001, $d$ = 2.91, $large$]. Accumulated head rotation also increased over time irrespective of the aid condition, which suggests that as participants found more gems over time, search became more difficult. Correspondingly, there was an overall effect of bin, $F$(1.27, 29.27) = 36.81, $p$ < 0.001, $\eta_{}^{\mathrm{2}}$ = .68, $large$, such that head rotation significantly increased over the course of each bin e.g. bin 1 versus bin 4, $t$(23) = -8.90, $p$ < 0.001, $d$ = -2.54, $large$. 

Based on visual inspection of Figure \ref{fig:rotation} it appears that accumulated head rotation was greater in bins 1-3 of the no aid condition compared to the other aid conditions, suggesting that the arrows and compass aided search in the first three bins. However, in the final bin participants also had greater accumulated head rotation in the arrow condition compared to both the radar and the compass, which suggests that when searching for the final 6 gems, the arrows condition no longer benefited search performance as total head rotation was increased. Consistent with this interpretation, there was a significant interaction ($F$(2.60,59.90) = 12.10, $p$ < 0.001, $\eta_{}^{\mathrm{2}}$ = .35, $large$). To break down the interaction, we examined the effect of bin at each level of the aid factor. There was an effect of bin in each of the navigation aid conditions- a similar pattern across each aid condition was found such that there was significantly less accumulated head rotation in earlier bins compared to later bins e.g., bin 1 versus bin 4 (all p's < 0.05). In order to further break down this interaction we conducted four separate ANOVA's for the effect of aid at each level of bin. In bins 1-3 accumulated head rotation was greater in the no aid condition compared to the arrows and compass condition (all p's < 0.05). In the final bin, participants had greater accumulated head rotation in the no-aid condition compared to the radar and the compass; no-aid versus radar, [$t$(23) = 6.21, $p$ < 0.001, $d$ = 1.79, $large$]; no-aid versus compass, [$t$(23) = 6.35, $p$ < 0.001, $d$ = 1.87, $large$]. In the final bin participants also had greater accumulated head rotation in the arrow condition compared to both the radar and the compass; arrow versus radar, [$t$(23) = 2.90, $p$ < 0.05, $d$ = 0.83, $large$]; arrow versus compass, [$t$(23) = 3.64, $p$ < 0.01, $d$ = 0.91, $large$].

\begin{figure}[t]
\centering \includegraphics[width = \columnwidth]{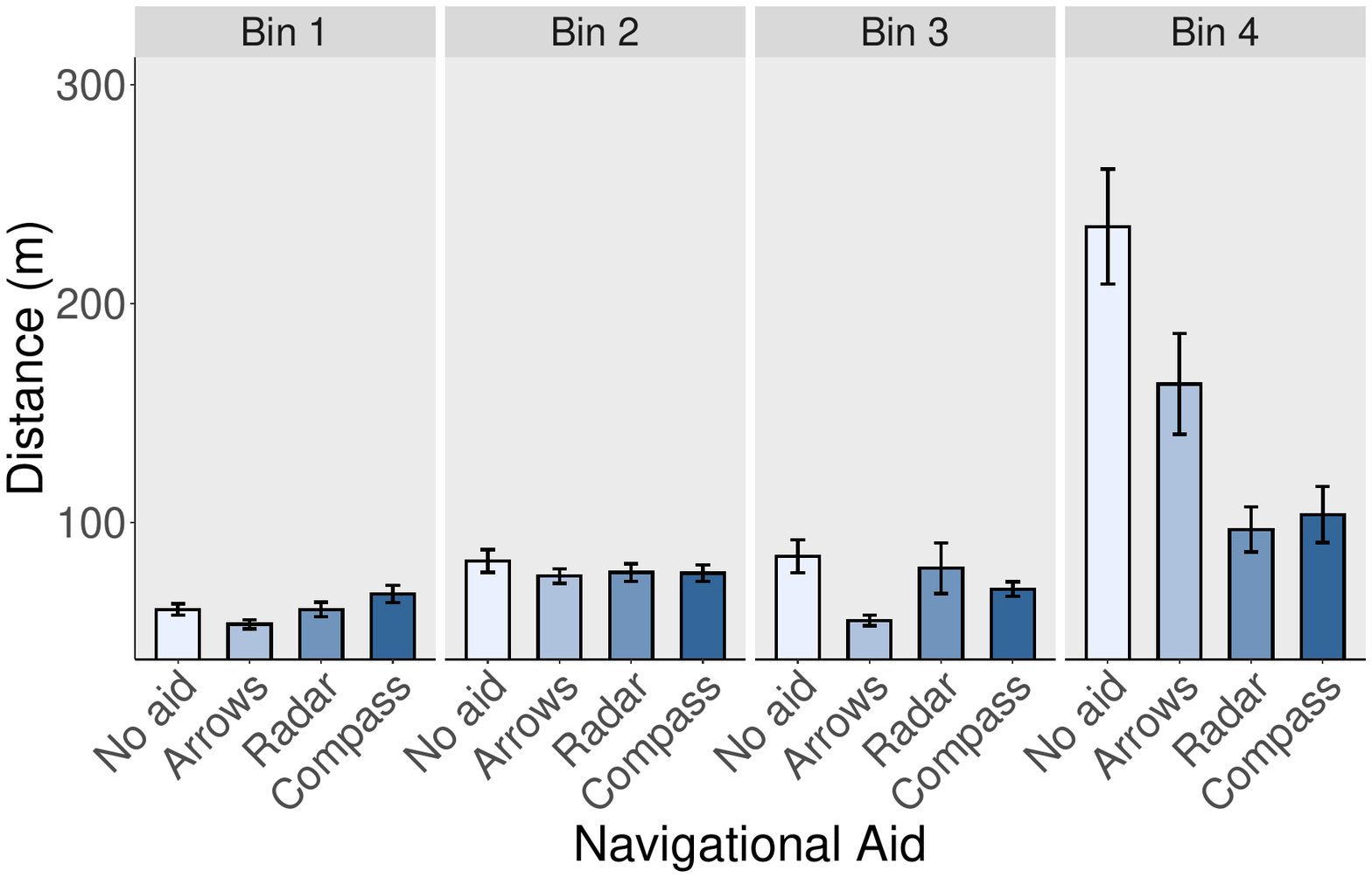}
\caption{Total distance traveled (in meters) plotted as a function of navigation aid and bin. Participants traveled a significantly longer distance in the no aid condition compared to the other three aid conditions. They also traveled a longer distances as fewer gems remained to be found.}
\Description{Total distance traveled (in meters) plotted as a function of navigation aid and bin. Participants traveled a significantly longer distance in the no aid condition compared to the other three aid conditions. They also traveled a longer distances as fewer gems remained to be found}
\label{fig:distance}
\end{figure}

\paragraph{Distance Traveled}

\begin{figure*}[t]
\centering
\includegraphics[width=\textwidth]{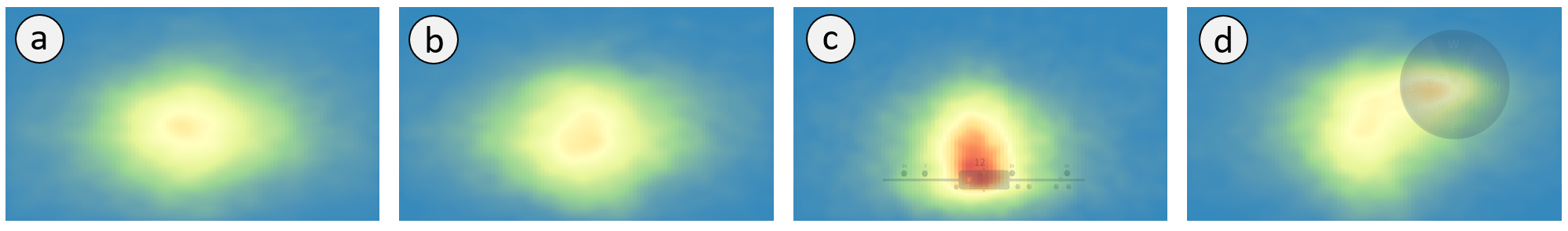}
\caption{Heat maps of the proportional density of gaze samples for each of the four aid conditions, accumulated over all participants with complete datasets: (a) No aid, (b) Arrows, (c) Compass and (d) Radar. The compass and radar areas-of-interest are also indicated in (c) and (d) respectively. }
\Description{Heat maps of the proportional density of gaze samples for each of the four aid conditions, accumulated over all participants with complete datasets: (a) No aid, (b) Arrows, (c) Compass and (d) Radar. The compass and radar areas-of-interest are also indicated in (c) and (d) respectively. Details are described in the text.}
% Users had a higher gaze density in the compass area of interest during the compass aid condition when compared to the other three conditions, and a higher gaze density in the radar area of interest during the radar aid condition when compared to the other three conditions.
\label{fig:heatmaps}
\end{figure*}

Distance traveled during the task is plotted as a function of navigation aid and bin in Figure \ref{fig:distance}. Overall, results revealed that the distance participants traveled was affected by aid ($F$(3,69) = 13.31, $p$ < 0.001, $\eta_{}^{\mathrm{2}}$ = .37, $large$). Participants traveled a significantly longer distance in the no aid condition compared to the other three aid conditions, suggesting that the aids benefited performance; no aid versus arrows, [$t$(23) = 3.39, $p$ < 0.05, $d$ = 0.95, $large$]; no aid versus radar, [$t$(23) = 5.47, $p$ < 0.001, $d$ = 1.43, $large$]; no aid versus compass conditions, [$t$(23) = 5.04, $p$ < 0.001, $d$ = 1.45, $large$]. There was also an overall effect of bin on distance traveled ($F$(1.26, 29.00) = 51.93, $p$ < 0.001, $\eta_{}^{\mathrm{2}}$ = .69, $large$) such that participants traveled a significantly longer distance over the course of each bin e.g., bin 1 vs. bin 4, $t$(23) = -8.73, $p$ < 0.001, $d$ = -1.90, $large$.

In addition to the overall effects of aid and bin, visual inspection of Figure \ref{fig:distance} demonstrates that the effect that the navigation aid had on distance traveled was modulated by bin. Participants traveled less distance in the arrows condition relative to the no-aid and compass in bins 1 and 3. However, in the final bin participants traveled a longer distance in the no aid and arrows condition relative to the both the screen-stabilized aids. Consistent with this interpretation, the interaction between aid and bin was significant ($F$(2.81,64.72) = 51.93 $p$ < 0.001, $\eta_{}^{\mathrm{2}}$ = .30, $large$). To break down the interaction we examined the effect of bin at each level of aid. There was an overall effect of bin in the no aid, arrows and radar conditions such that participants traveled a significantly shorter distance in the earlier compared to later bins in both of these conditions e.g., bin 1 vs. bin 4 (all p's < 0.05). There was also an effect of bin in the compass condition but none of the pairwise comparisons survived correction. To further break down the interaction we also conducted separate ANOVAs for the effect of aid at each level bin. There was an effect of aid in bin 1 ($F$(3,69) = 4.03 $p$ < 0.05, $\eta_{}^{\mathrm{2}}$ = .15, $large$) such that participants traveled significantly less distance in the arrows than the compass condition, [$t$(23) = -4.07, $p$ < 0.01, $d$ = -0.91, $large$]. Navigation aids were not found to have an impact on distance traveled in bins 2 and 3. However, there was an effect in bin 4 ($F$(2.13,49.12) = 12.19, $p$ < 0.001, $\eta_{}^{\mathrm{2}}$ = .35, $large$) such that participants traveled a significantly longer distance in the no-aid condition compared to the radar and the compass condition; no aid versus radar [$t$(23) = 5.10, $p$ < 0.001, $d$ = 1.42, $large$]; no aid versus compass [$t$(23) = 4.56, $p$ < 0.001, $d$ = 1.30, $large$]. Participants no longer traveled a shorter distance in the arrows condition relative to the no-aid condition when searching for the final 6 gems.

\subsection{Eye Gaze}

Eye gaze was recorded using the Hololens-2's built in eye-tracking hardware. Of the 24 participants, 3 had incomplete gaze datasets for one or more trials and hence were excluded from all eye gaze analyses. We examined whether participants utilized the head-stabilized aids during search, and also analyzed the impact of the head-stabilized aids on user attention to the screen and the environment.

% Although eye gaze data is collected at 30fps by the headset, our app recorded data at 5fps due to bandwidth and performance constraints. 

% \paragraph{Preliminary Analysis}
% To assess whether participants' eyes were drawn to the radar, which was on the right side of the display we compared the mean horizontal (\emph{x}) position in the no-aid compared to the horizontal position in the radar condition. As the compass was in the bottom portion of the screen, we compared the vertical (\emph{y}) position in the no aid condition compared to the compass condition. Two paired samples t-tests were used to assess these comparisons. The eye gaze $y$-position was significantly smaller in the compass condition ($M$ = -0.26) compared to the no aid condition ($M$ = -0.32), suggesting that participants eye-gaze was overall lower in the display likely due to more time spent looking at the compass ($t$(20) = 4.99, $p$ < 0.001, $d$ = 0.84, $large$). However, there was no difference in the eye gaze $x$-position between the radar ($M$ = 0.045) and no aid condition ($M$ = 0.064), suggesting that participants may not have fully made use of the radar to aid search ($t$(20) = 0.44, $p$ = 0.67, $d$ = 0.12). 

\begin{figure}[t]
% \centering \includegraphics[width=6.8cm, keepaspectratio]{img/audioAcc.png}
\begin{subfigure}[b]{0.45\columnwidth}
         \includegraphics[width=\columnwidth]{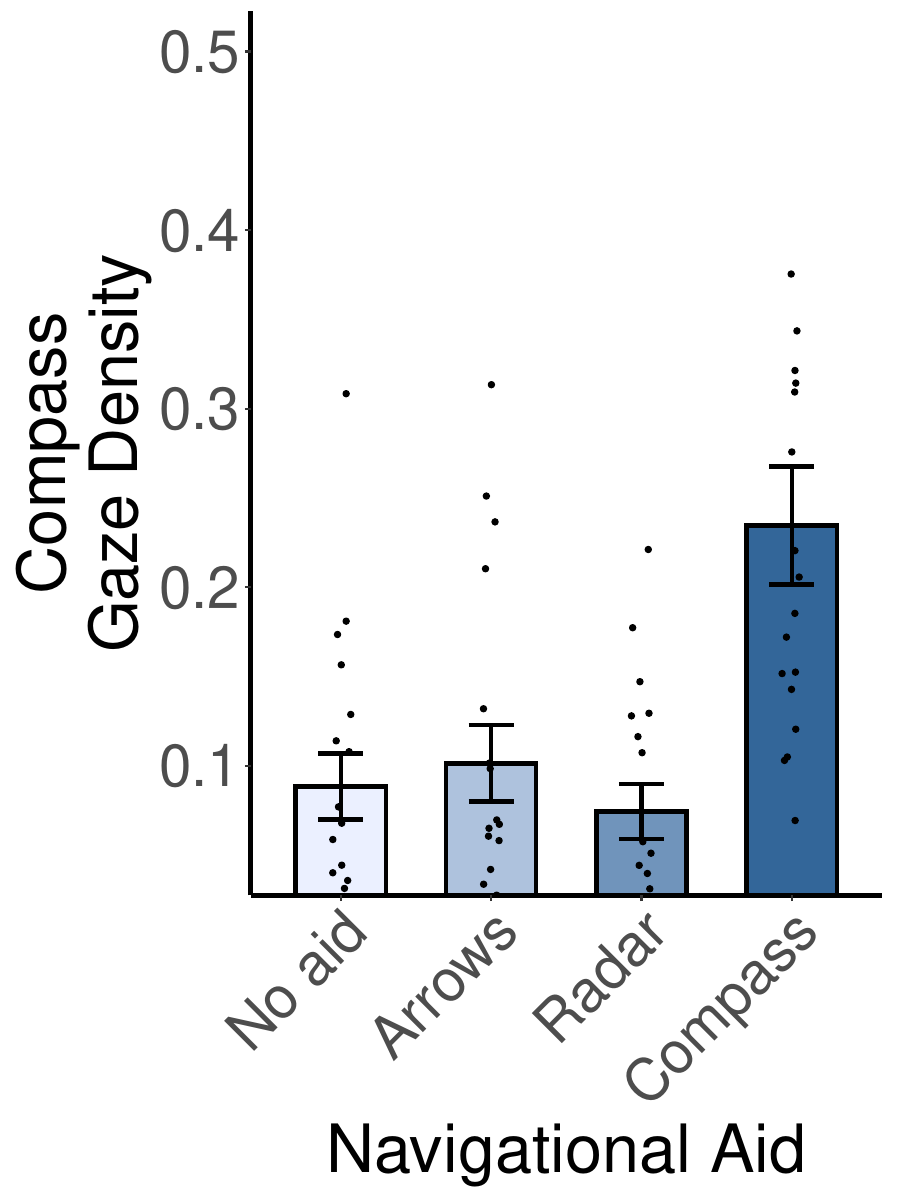}
     \end{subfigure}
     \hfill
     \begin{subfigure}[b]{0.45\columnwidth}
         \includegraphics[width=\columnwidth]{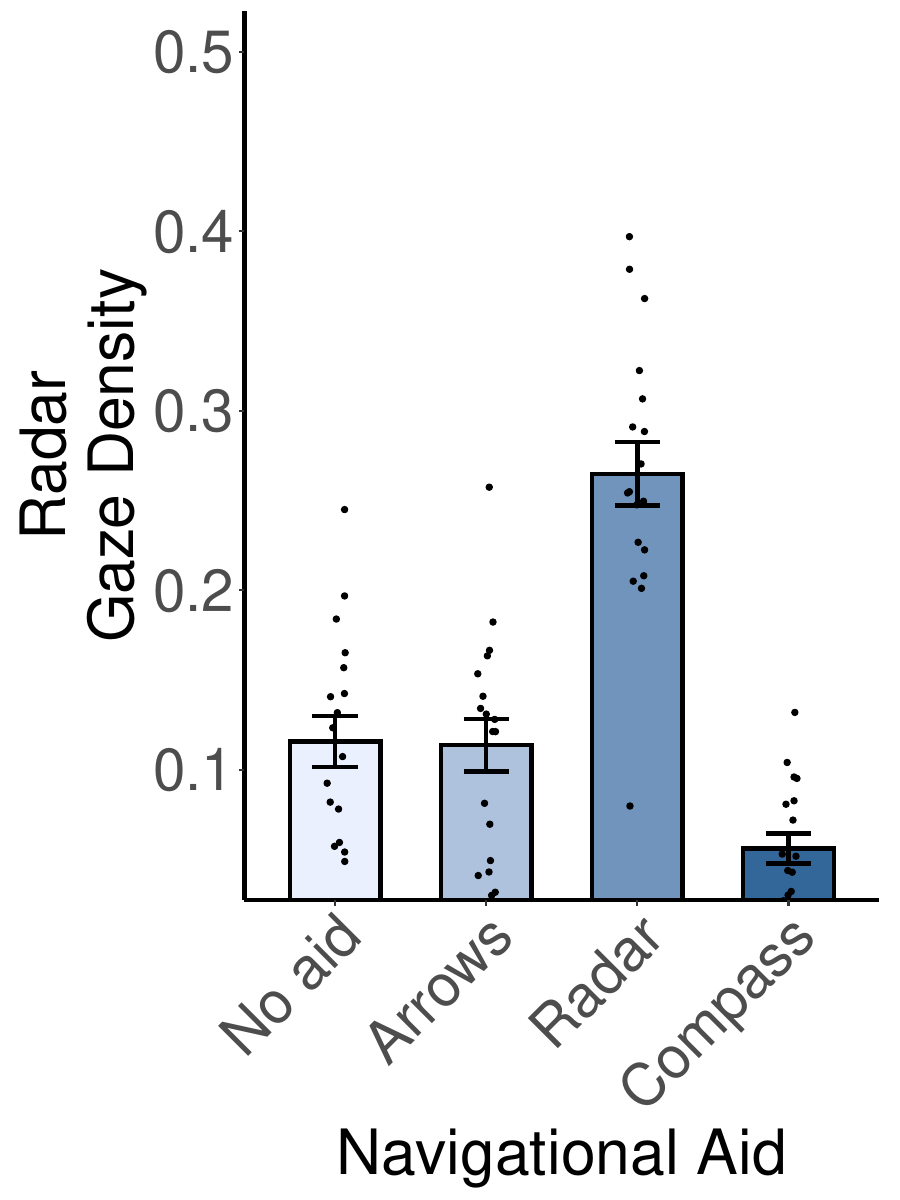}
     \end{subfigure}
\caption{Density of gaze (proportion of gaze samples) in the two areas of interest: compass (left) and radar (right) as function of navigation aid. Gaze density was significantly higher in the compass condition compared to the other three for the compass area of interest, and significantly higher in the radar condition compared to the other three for the radar area of interest. For this and all following figures: Black dots represent individual data points.}
\Description{Plot of the density of gaze (proportion of gaze samples) in the two areas of interest: compass (left) and radar (right) as function of navigation aid. Gaze density was significantly higher in the compass condition compared to the other three for the compass area of interest, and significantly higher in the radar condition compared to the other three for the radar area of interest.}
\label{fig:gazeDensity}
\end{figure}

\begin{figure}[t]
% \centering \includegraphics[width=6.8cm, keepaspectratio]{img/audioAcc.png}
\begin{subfigure}[b]{0.45\columnwidth}
         \includegraphics[width=\columnwidth]{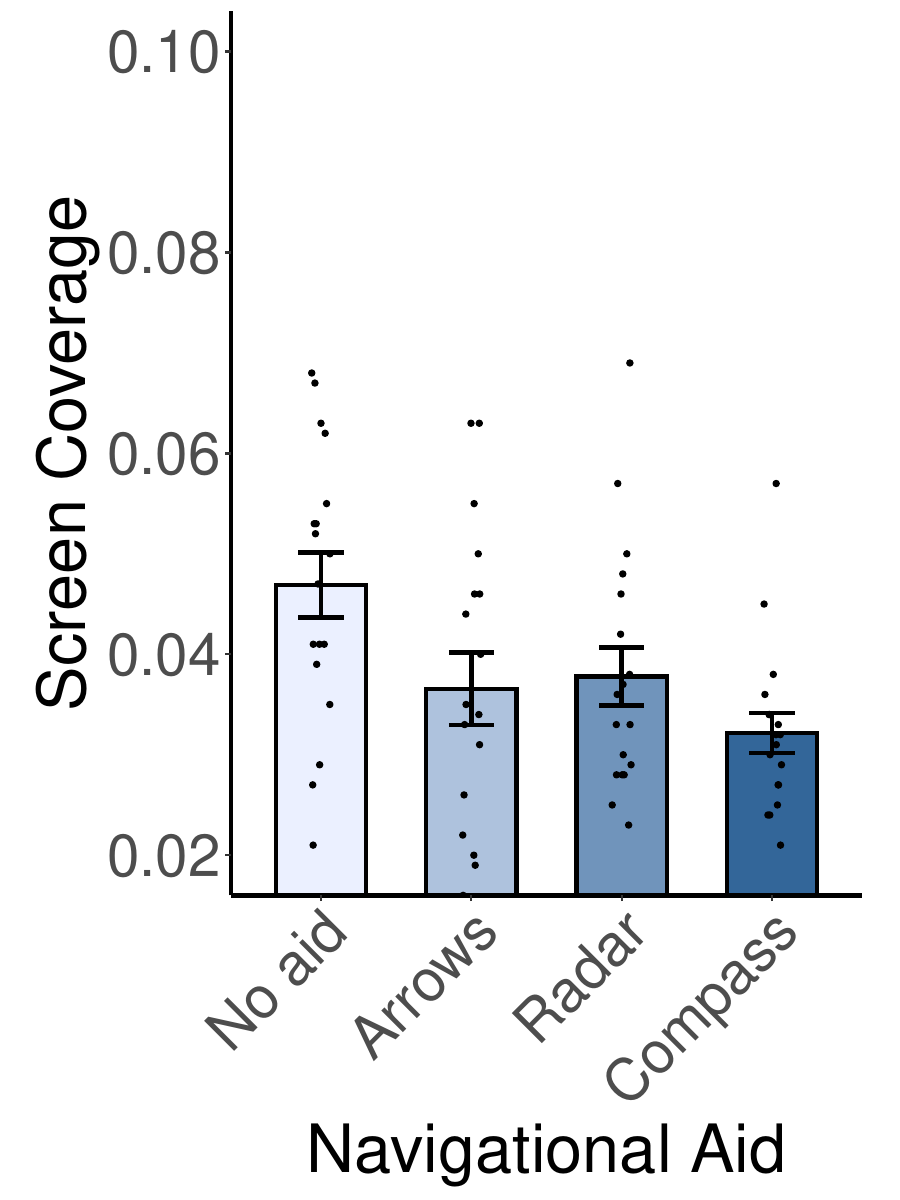}
     \end{subfigure}
     \hfill
     \begin{subfigure}[b]{0.45\columnwidth}
         \includegraphics[width=\columnwidth]{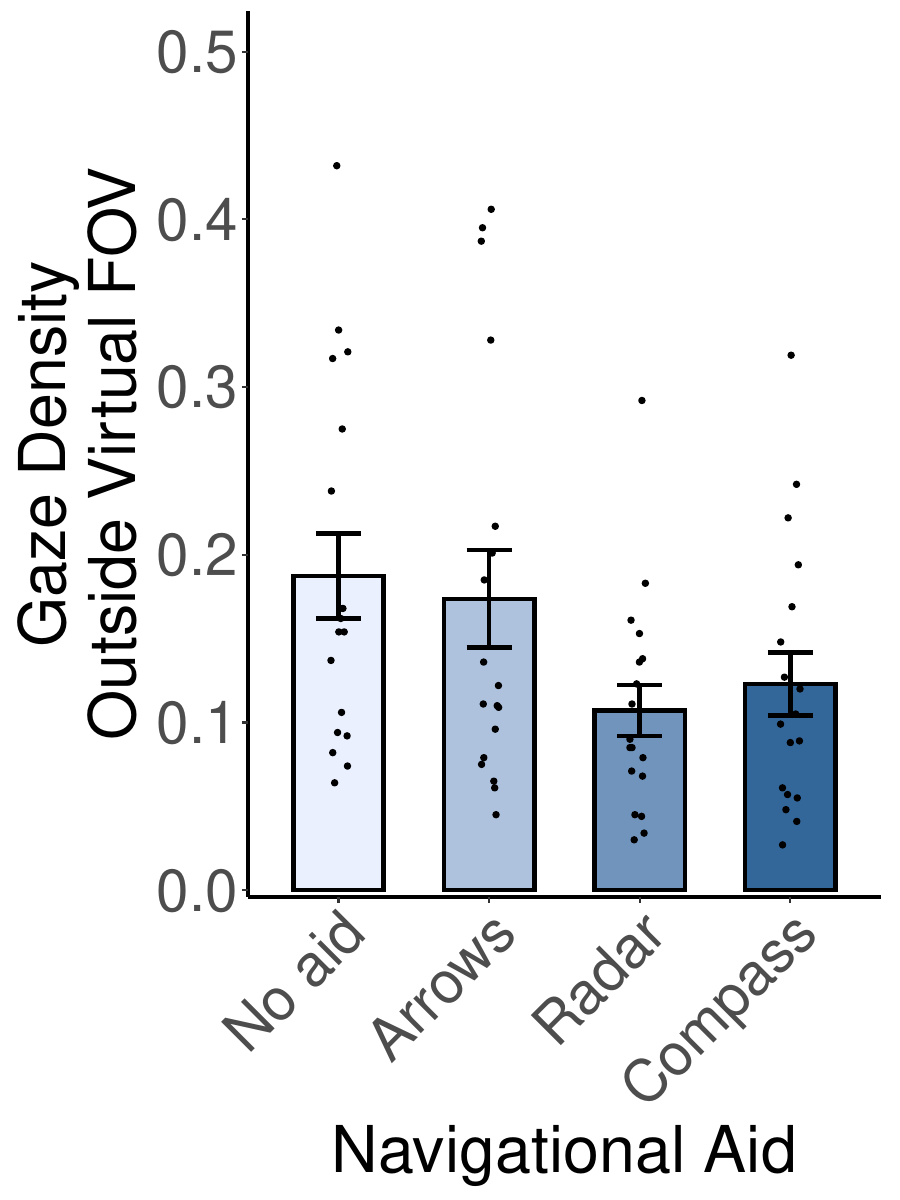}
     \end{subfigure}
\caption{Left: Screen coverage (proportion of pixels on the screen that were viewed at least once), as a function of navigation aid. Screen coverage for the compass condition was significantly less than the no aid condition. Right: Proportion of gaze samples outside the virtual field of view (FOV), as a function of navigation aid. A significantly larger proportion of samples were outside the field of view in the no aid condition compared to the compass and radar conditions, suggesting that users paid more attention to the screen (and hence less attention to the physical environment) in the latter two conditions.}
\Description{Left: Screen coverage (proportion of pixels on the screen that were viewed at least once), as a function of navigation aid. Screen coverage for the compass condition was significantly less than the no aid condition. Right: Proportion of gaze samples outside the virtual field of view (FOV), as a function of navigation aid. A significantly larger proportion of samples were outside the field of view in the no aid condition compared to the compass and radar conditions, suggesting that users paid more attention to the screen (and hence less attention to the physical environment) in the latter two conditions.}
\label{fig:screenCoverage}
\end{figure}

\paragraph{Correction of Temporal Offset}

Visual inspection of the gaze data in relation to user movement, using the playback software,  showed that the reported "gaze origin" often trailed behind the user's head position. The presence of a systematic spatial offset of the gaze target has been reported in the literature \cite{aziz2022assessment}, and our observations suggest that there is also a temporal offset of recorded gaze origin and target which is likely highlighted during a state of motion. 
% This offset was apparent even relative to the head position (which was recorded at the same frequency). We therefore concluded that the offset was not an artefact of the frequency of recording, and instead originated in the built-in eye tracking hardware/software.

A post-hoc recalibration of the data would require controlled gaze information for each participant, which had not been collected during the experiment. We therefore optimized the Euclidean distance between the head position and reported gaze origin during each trial, and found that advancing the gaze data by 5 samples (approximately 1s) led to minimum offset between the two points. During this process 3 participants, who had one or more trials with a normalized Euclidean offset larger than 5 times the average, were excluded from further analysis.

\paragraph{Area of Interest Analysis}

Visual inspection of Figure \ref{fig:heatmaps} suggests that users had a higher gaze density (proportion of gaze samples) in the compass area of interest during the compass aid condition (Figure \ref{fig:heatmaps}c) when compared to the other three conditions and a higher gaze density in the radar area of interest during the radar aid condition (Figure \ref{fig:heatmaps}d) when compared to the other three conditions, which confirms that participants utilized the head-stabilized tools in their search. An area-of-interest analysis comparing the density of gaze samples in each area of interest on the screen - the radar and the compass - was performed to examine the impact of the two head-stabilized aids on user focus. Results are shown in Figure \ref{fig:gazeDensity}.

There was a main effect of aid on the density of gaze in the compass area of interest ($F$(1.55,26.3) = 26.728, $p$ < 0.001, $\eta_{}^{\mathrm{2}}$ = .61, $large$), with significantly higher gaze density in the compass aid condition when compared to the other three conditions; no aid versus compass [$t$(18) = -6.85, $p$ < 0.001, $d$ = -1.29, $large$]; compass versus radar [$t$(18) = 5.43, $p$ = 0.001, $d$ = 1.47, $large$], compass versus arrows [$t$(18) = 5.34, $p$ < 0.001, $d$ = 1.13, $large$]. There was also a main effect of aid on the density of gaze in the radar area of interest ($F$(1.93,32.73) = 55.983, $p$ < 0.001, $\eta_{}^{\mathrm{2}}$ = .77, $large$), with significantly higher gaze density in the radar aid condition when compared to the other three conditions; no aid versus radar [$t$(18) = -7.05, $p$ < 0.001, $d$ = -2.20, $large$]; compass versus radar [$t$(18) = -10.72, $p$ = 0.001, $d$ = -3.55, $large$], radar versus arrows [$t$(18) = 6.99, $p$ < 0.001, $d$ = 2.20, $large$].

\paragraph{Screen Coverage}

Screen coverage was computed as the proportion of pixels on the virtual display that were viewed at least once, and is shown in Figure \ref{fig:screenCoverage}, on the left. An ANOVA with aid condition as a within-subjects variable highlighted a main effect of aid on the screen coverage($F$(3,51) = 5.91, $p$ = 0.002, $\eta_{}^{\mathrm{2}}$ = .26, $large$), with significantly higher screen coverage in the no aid condition when compared to the compass [$t$(18) = 4.15, $p$ = 0.004, $d$ = 1.29, $large$].

An analysis of the proportion of gaze samples outside the virtual field-of-view (shown in Figure \ref{fig:screenCoverage} on the right) suggested that participants paid more attention to the screen in the two head-stabilized conditions (compass and radar) when compared to the no aid condition, as evidenced by fewer gaze samples outside the virtual field of view. There was a main effect of aid on the gaze density outside the virtual field of view ($F$(2.32,39.38) = 7.58, $p$ = 0.001, $\eta_{}^{\mathrm{2}}$ = .31, $large$), with significantly more samples outside the virtual field of view in the no aid condition when compared to the compass [$t$(18) = 3.05, $p$ = 0.043, $d$ = 0.68, $large$] and radar [$t$(18) = 4.58, $p$ = 0.002, $d$ = 0.91, $large$].

These results indicate reduced attention to the physical world in the presence of head-stabilized aids, which could partially explain the results of the object recall task (Section \ref{sec:recall}).

\subsection{Audio Response}

\begin{figure}[t]
% \centering \includegraphics[width=6.8cm, keepaspectratio]{img/audioAcc.png}
\begin{subfigure}[b]{0.45\columnwidth}
         \includegraphics[width=\columnwidth]{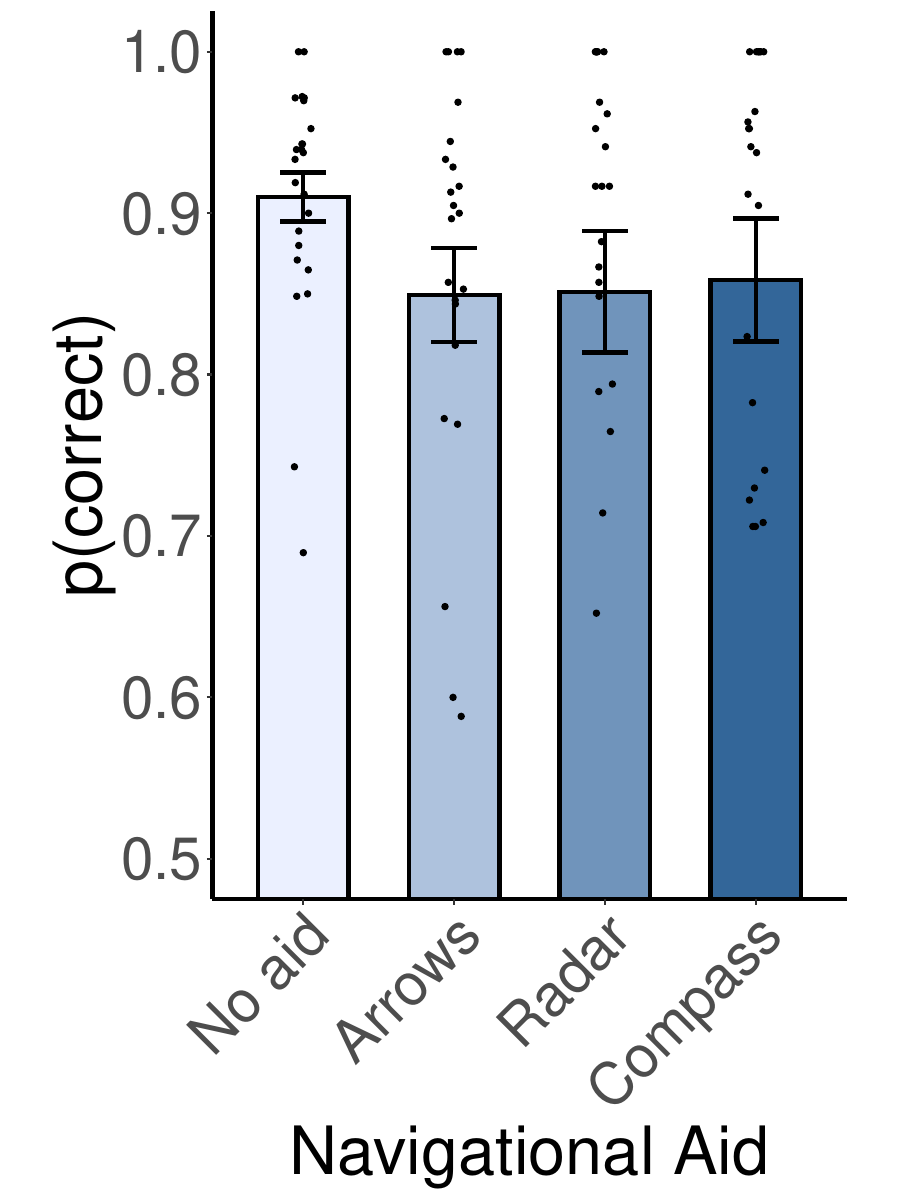}
     \end{subfigure}
     \hfill
     \begin{subfigure}[b]{0.45\columnwidth}
         \includegraphics[width=\columnwidth]{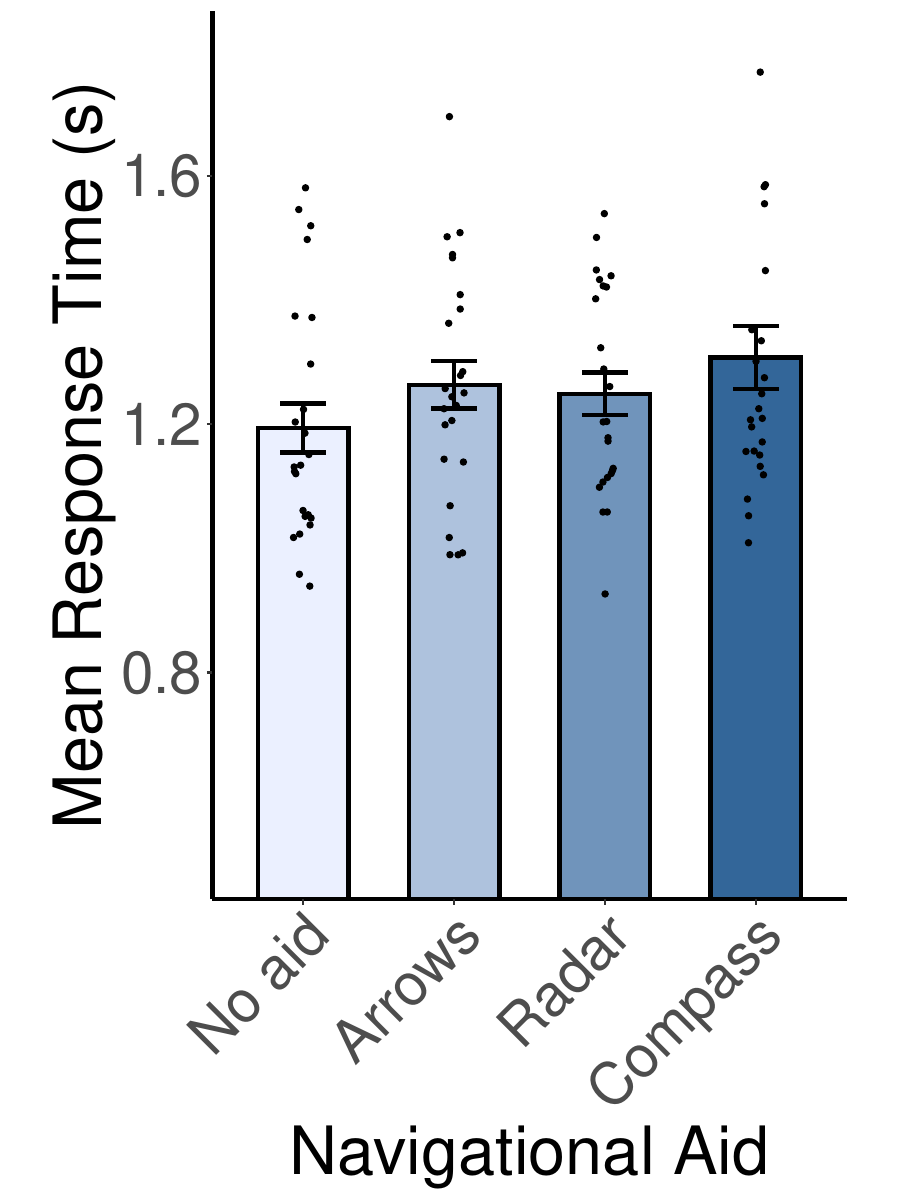}
     \end{subfigure}
\caption{Mean accuracy (left) and mean response time (right) on the audio task as function of navigation aid. Accuracy was higher with no aid when compared to any of the three navigation aids, and response time was significantly higher with the compass when compared to no aid.}
\Description{Mean accuracy (left) and mean response time (right) on the audio task as function of navigation aid. Accuracy was higher with no aid when compared to any of the three navigation aids, and response time was significantly higher with the compass when compared to no aid.}
\label{fig:audioAcc}
\end{figure}

Accuracy and mean response time to the audio task are plotted as a function of navigation aid condition in Figure \ref{fig:audioAcc}. Findings indicated that the addition of a navigation aid, in particular the compass, reduced performance in the audio control-task. The effect of navigation aid condition on audio accuracy approached significance such that accuracy was qualitatively higher in the no aid condition compared to the other three aid conditions ($F$(3,69) = 2.18, $p$ = 0.099, $\eta_{}^{\mathrm{2}}$ = .086, $medium$). The effect of navigation aid on audio task reaction time was significant ($F$(2.12,48.74) = 3.93, $p$ < 0.05, $\eta_{}^{\mathrm{2}}$ = .15, $large$), such that participants were significantly slower in the compass condition than the no aid condition [$t$(23) = -3.56, $p$ = 0.01, $d$ = -0.51, $medium$].

\subsection{User Preferences}

\begin{figure}[t]
\centering \includegraphics[width=0.85\columnwidth, keepaspectratio]{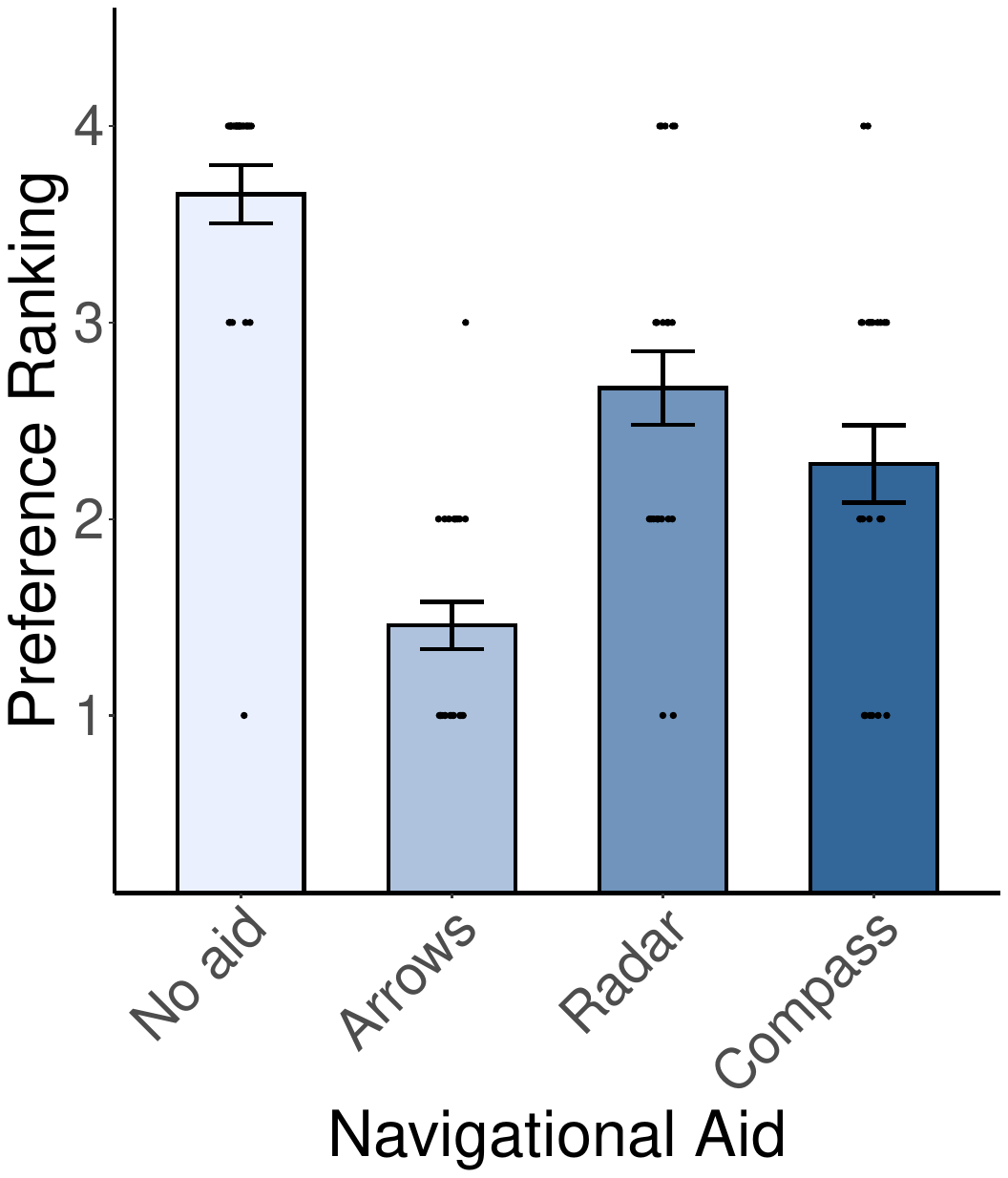}
\caption{Mean preference ranking for each of the navigation aid conditions as a function of navigation aid, a lower value indicates a higher rank. The arrows aid was the most preferred, and no aid least preferred.}
\Description{Mean preference ranking for each of the navigation aid conditions as a function of navigation aid, a lower value indicates a higher rank. The arrows aid was the most preferred, and no aid least preferred.}
\label{fig:preferenceRanking}
\end{figure}

Mean preference ranking for each navigation aid is shown in Figure \ref{fig:preferenceRanking}, with lower values indicating higher preference ranking (1-4, 1 being most preferred). The arrows aid was the most preferred (and no aid least preferred), as indicated by a Friedman's test ($\chi_{}^{\mathrm{2}}$(3) = 34.67, $p$ < 0.001, $effsize$ = .50, $medium$). Pairwise comparisons using the Wilcoxon test revealed that there was a preference for the arrow aid compared to the no aid [$W$ = 0, p < 0.001, $effsize$ = 0.90, $large$], radar [$W$ = 0, p < 0.001, $effsize$ = 0.90, $large$], and compass [$W$ = 0, p < 0.001, $effsize$ = 0.84, $large$]. There was also a preference for the compass over no aid [$W$ = 0, p < 0.001, $effsize$ = 0.87, $large$], and radar over no aid [$W$ = 171, p < 0.001, $effsize$ = 0.84, $large$].

\subsection{Object Recall}
\label{sec:recall}
\begin{figure}[t]
\centering \includegraphics[width=\columnwidth, keepaspectratio]{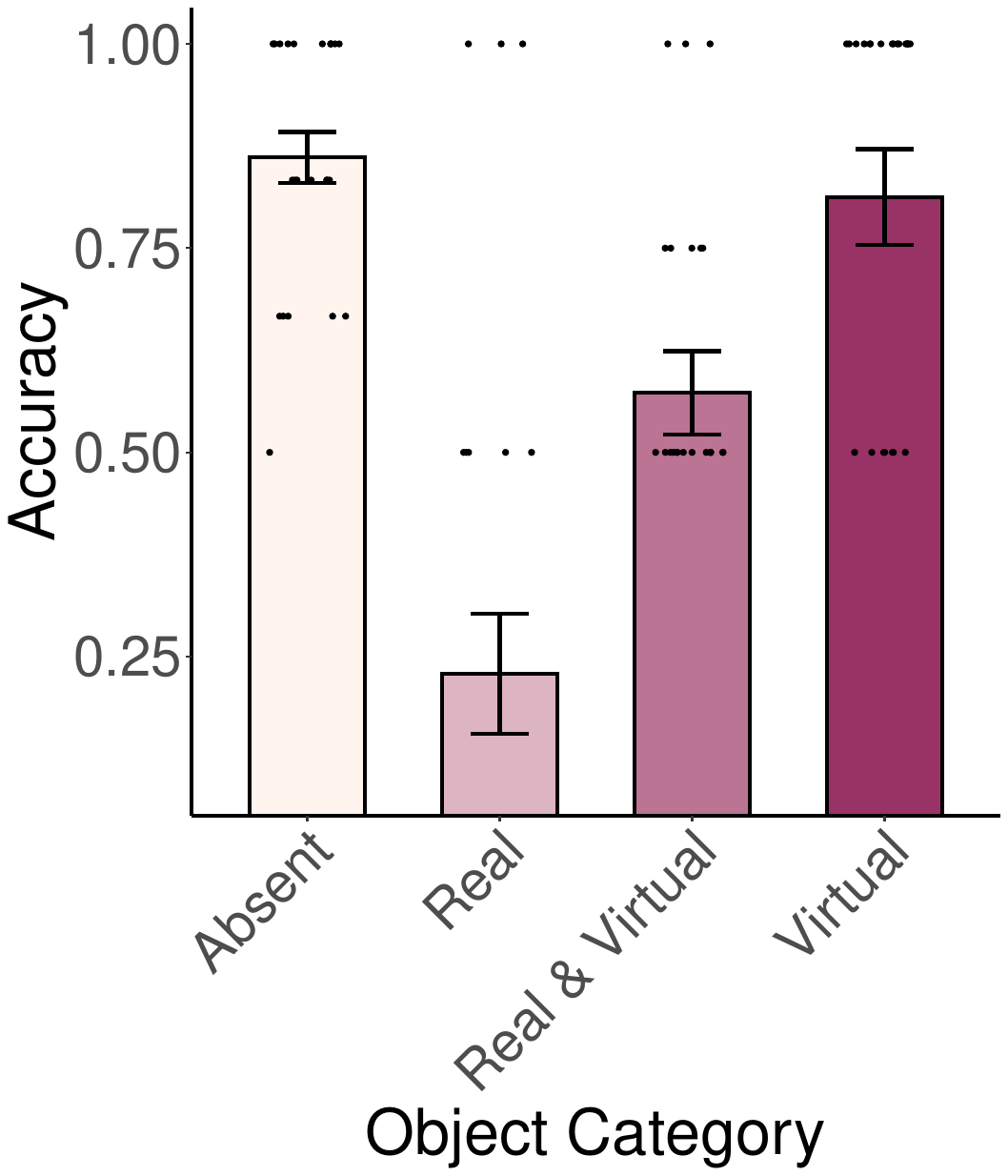}
\caption{Mean accuracy of responses to object recall questions, as a function of object category. Participants were significantly less accurate at classifying real objects than any of the other object categories.}
\Description{Mean accuracy of responses to object recall questions, as a function of object category. Participants were significantly less accurate at classifying real objects than any of the other object categories.}
\label{fig:objectCategory}
\end{figure}

 Recall accuracy for each category of objects is shown in Figure  \ref{fig:objectCategory}. Participants were significantly less accurate at classifying real objects than all other object categories, suggesting that participants were least aware of the physical objects in the environment. A one-way repeated measures ANOVA with the factor object type (absent, real, real and virtual, virtual) was conducted on classification accuracy. There was a main effect of object category ($F$(3,69) = 29.85, $p$ < 0.001, $\eta_{}^{\mathrm{2}}$ = .57, $large$). Pairwise comparisons revealed that participants were significantly less accurate at classifying real objects than virtual objects [$t$(23) = -6.23, $p$ < 0.001, $d$ = -1.79, large], both virtual and real objects [$t$(23) = -8.17, $p$ < 0.001, $d$ = -1.11, $large$], and objects that were absent [$t$(47) = 7.75, $p$ < 0.001, $d$ = 2.28, large].
 
\section{Discussion}

All three aids were found to improve performance on the gem search task relative to the no-aid condition as indicated by reduced total session time and head rotation. There was evidence that the in-world arrows had a slight benefit over the on-screen aids over the first three bins in a trial as there was a comparatively lower total session time \blue{and higher walking speed} in this condition, and this was also reflected in user feedback which demonstrated a strong preference for the in-world arrows over the on-screen aids. Between the two on-screen aids there was a trend to better performance with the compass, which was again supported by a trend to user preference for the compass over the radar.

The impact of augmented reality technology and navigation aids on user attention is an important consideration in the design of AR applications. There is a possible cognitive cost associated with the use of some of these aids in AR that can negatively impact multitasking ability, as evidenced by poorer performance on the audio response task when using the compass compared to the control condition. The compass condition may have introduced a higher cognitive load compared to other conditions, which resulted in a performance impairment in the secondary audio task. It is well documented that when simultaneously attending to two challenging tasks from different modalities, such as visual and auditory, the performance in one or both tasks will be reduced \cite{spence1997audiovisual}. This is an effect that has been extensively reported, e.g., regarding driver distraction while using a cell phone, where limited attentional resources are directed towards auditory information, and driving performance may be impaired \cite{strayer2001driven}. Therefore, there can be some safety concerns when introducing AR navigation aids, depending on the task. If the UI is complex, then users may not have sufficient attentional resources available for completing other tasks. 

Furthermore, situational awareness of the physical environment was also found to be impacted, as there was reduced recall for physical objects compared to all other object categories. Such an effect has been demonstrated previously \cite{kim2022investigating} but with the introduction of the navigation aids, this effect was found to be considerably larger (Cohen's d = 1.37 vs 1.79). The results of the eye gaze analysis suggest that the presence of the navigation aids partially explains this effect, with a higher proportion of gaze samples within the virtual field of view and reduced screen coverage in the presence of the compass and radar. This is in line with research demonstrating that distraction reduces situational awareness of the environment e.g., \cite{kass2007effects}, and could have important implications for the design of augmented reality applications for wide-area use. If it is desired that users form and recall a useful spatial model of the physical environment, then designers and application programmers may want to emphasize (for example by highlighting in AR) important physical objects instead of just adding virtual ones. And they should have an eye on avoiding unnecessary virtual clutter and distraction. 

Although the in-world arrows did demonstrate some quantitative advantages over the on-screen aids, \blue{including a faster walking speed}, the evidence was not as conclusive as we originally expected and also did not fully mirror the clear user preference for them. While attempting to find and collect the last quarter of the hidden gems (during bin 4), the arrow condition exhibited disadvantages \blue{in task time} compared to the on-screen conditions, which more easily facilitated a process of simply pursuing the next uncollected gem, or even of planning and refining path optimizations. Arrows also had a bit of an unfair disadvantage in bin 4, as the arrows for some (up to 3) gems in the scene were not always visible but possibly partially hidden by physical infrastructure such as columns and building corners. In contrast, gem representations were always visible on-screen for all remaining gems in the compass and radar conditions. One could certainly improve the arrow aid by depicting arrows even when occluded, using some form of X-ray vision visualization \cite{avery2009improving,zollmann2010image}, perhaps in a different rendering style that indicates that the arrow is currently located behind a physical occluder (\cite{livingston2003resolving}).  

In-world navigation aids require very precise tracking capabilities for AR registration (which is still a challenge in outdoors wide-area environments even with the current state-of-the-art technology), whereas on-screen aids in absence of world-stabilized AR could enjoy much more leeway with regard to user pose tracking accuracy. Therefore, our results suggest that there may well be still a place and time for on-screen navigation aids such as the compass and radar. It is interesting that Microsoft's SDK recommendations actively discourage head-up display components~\cite{tieto2022hololens} (see also limitations section below) given that they could clearly still be useful in certain situations (depending on the application), especially in tracking-challenged head-mounted wide-area augmented reality scenarios.
\vspace*{-1ex}

\subsection{Limitations}
The present study had several limitations. First, registration errors did occur in a few trials, which were abandoned and repeated with a backup layout instead. This could have introduced some variability into participants' experience, though experimenters did their best to ensure minimal disruption to the procedure in those situations.
Secondly, although there was some pedestrian traffic in the experiment area due to the experiment being conducted in a public space, experimenters ensured that participants were minimally impacted (if at all) by directing all traffic away from the participants' location.
Third, there was some jitter of the on-screen aids during user movement because these aids were head-stabilized (and hence tethered to head motion). Even though head-stabilized content is not recommended in head-mounted displays for this reason~\cite{tieto2022hololens}, participant ratings of three elements of the Simulator Sickness Questionnaire~\cite{kennedy1993simulator} (namely headache, eyestrain and blurred vision) didn't reveal any difference in discomfort levels between on-screen and in-world aids (all $p$ values > 0.05). The effect of this jitter could be further investigated by comparing head stabilized navigation aids with those that use body stabilization, and hence allow content to follow the user with relatively smooth motion.

Further, an analysis of eye gaze data revealed a temporal offset of the gaze information even though the headset was calibrated for each participant, which was likely accentuated by the quick movements often required in the task. This could not be corrected completely in the absence of a controlled recalibration task for each participant. The inclusion of such a task in future experiments will improve the quality of corrected gaze data.

\section{Conclusion}
Many real-world scenarios involve people searching within their surrounding environments, e.g. tourism, shopping, search and rescue operations. Augmented reality technology can support search tasks in wide-area environments, and the current work discusses the potential, opportunities, and implications of using this technology for navigation guidance. Potential side effects to be considered when designing AR applications for outdoor use were discussed and some design recommendations derived. Specifically, this paper presented a wide-area outdoor user study examining the impact of navigation aids on user search performance, and also examined spatial awareness of the environment during this task. Regarding our first question, there was a strong user preference for world-stabilized in-world annotations when compared to head-stabilized on-screen ones, accompanied also by some quantitative benefits. Controlled wide-area outdoor AR user studies are still few and far between, and establishing concrete benefits of direct-overlay registered AR is a significant finding for the AR community. 

At the same time, performance of the screen-stabilized aid conditions in the search and response tasks did not reflect any major disadvantages either. This suggests that on-screen aids could still be useful, especially in situations where the objects in the environment are likely to change position (e.g. aftermath of an earthquake) and tracking accuracy may be impaired. The presence of some virtual annotations also reduced multitasking ability, and there was a general lack of attention to physical objects in the environment during all the tasks, answering our second question. This highlights important design considerations that must be taken into account when creating virtual content for outdoor augmented reality, such as potentially highlighting physical objects and keeping virtual clutter low. The controlled inclusion of physical search targets in addition to virtual ones in future experiments could help better understanding of these effects.

\begin{acks}
\red{This work was funded by the U.S. Army Combat Capabilities Development Command Soldier Center Measuring and Advancing Soldier Tactical Readiness and Effectiveness (MASTR-E) program through award W911NF-19-F-0018 under contract W911NF-19-D-0001 for the Institute for Collaborative Biotechnologies. Additional support came from ONR awards N00014-19-1-2553, N00014-20-1-2719, and N00014-23-1-2118, as well as NSF award IIS-1911230. The authors thank Alejandro Aponte and Emily Machniak for their assistance with data collection.}

% 
% Funding Body	Award Number	 
% U.S. Army Combat Capabilities Development Command Soldier Center	W911NF-19-F-0018	
% National Science Foundation	IIS-1911230	
% Office of Naval Research	 N00014-20-1-2719, N00014-19-1-2553	 

% 
\end{acks}

\bibliographystyle{ACM-Reference-Format}
\bibliography{references}

\appendix
\red{
\section{Additional Results - Lighting}
\label{sec:appendix-light}

While lighting was initially a factor in our analyses, distance traveled was the only dependent variable found to be affected by this factor. Further, lighting was not found to interact with any of the navigation aid conditions across our dependent measures. We therefore did not include lighting as a factor in the analyses reported in the paper, and discuss the results of the analyses from the paper with the additional factor of lighting condition (Natural and Night) here. 

A repeated measures ANOVA with the factors: lighting (natural, night), navigation aid (none, arrows, radar, and compass), and gem type (floating, physical, virtual) demonstrated that there was no main effect of lighting on discrimination accuracy $F$(1,22) = 0.42, $p$ = .52, $\eta_{}^{\mathrm{2}}$ = .019. There was also no interaction between lighting and aid ($F$(3,66) = 0.89, $p$ = 0.45, $\eta_{}^{\mathrm{2}}$ = .039), lighting and gem type ($F$(2,44) = 0.091, $p$ = 0.91, $\eta_{}^{\mathrm{2}}$ = .004) or a three-way interaction ($F$(3.78, 83.08) = 1.47, $p$ = 0.22 $\eta_{}^{\mathrm{2}}$ = .062). 

A repeated measures ANOVA with the factors: lighting, navigation aid, and bin (1, 2, 3, 4) was conducted on the three global metric variables total session time, head rotation, and distance traveled. 

The ANOVA examining the effects of these factors on total session time revealed that there was no main effect of lighting condition $F$(1,22) = 2.61, $p$ = 0.12, $\eta_{}^{\mathrm{2}}$ = .11. There was also no interaction between lighting and aid ($F$(3,66) = 1.39, $p$ = 0.25, $\eta_{}^{\mathrm{2}}$ = .059), lighting and bin ($F$(1.33, 29.32) = 3.67, $p$ = .054, $\eta_{}^{\mathrm{2}}$ = .14) or a three-way interaction ($F$(3, 65.99) = 1.65, $p$ = 0.19, $\eta_{}^{\mathrm{2}}$ = .070).

The ANOVA on head rotation revealed that there was no main effect of lighting condition ($F$(1,22) = 0.091, $p$ = 0.77, $\eta_{}^{\mathrm{2}}$ = .004). There was also no interaction between lighting and aid ($F$(3,66) = 0.78, $p$ = 0.51, $\eta_{}^{\mathrm{2}}$ = .034), lighting and bin ($F$(3, 66) = 0.51, $p$ = 0.68, $\eta_{}^{\mathrm{2}}$ = .023) or a three-way interaction ($F$(9, 198) = 0.57, $p$ = 0.82, $\eta_{}^{\mathrm{2}}$ = .025).

The ANOVA on speed revealed that there was no main effect of lighting condition ($F$(1,22) = 1.81, $p$ = 0.19, $\eta_{}^{\mathrm{2}}$ = .076). There was also no interaction between lighting and aid ($F$(3,66) = 0.42, $p$ = 0.74, $\eta_{}^{\mathrm{2}}$ = .019), lighting and bin ($F$(2.11, 46.4) = 0.11, $p$ = 0.91, $\eta_{}^{\mathrm{2}}$ = .005) or a three-way interaction ($F$(9, 198) = 0.79, $p$ = 0.63, $\eta_{}^{\mathrm{2}}$ = .035).

The ANOVA on accumulated head rotation revealed that there was no main effect of lighting condition ($F$(1,22) = 1.88, $p$ = 0.18, $\eta_{}^{\mathrm{2}}$ = .079). There was also no interaction between lighting and aid ($F$(3,66) = 1.92, $p$ = 0.14, $\eta_{}^{\mathrm{2}}$ = .08), lighting and bin ($F$(1.32, 28.95) = 3.49, $p$ = 0.061, $\eta_{}^{\mathrm{2}}$ = .14) or a three-way interaction ($F$(2.81, 61.80) = 2.52, $p$ = 0.095, $\eta_{}^{\mathrm{2}}$ = .093).

\begin{figure}[t]
\centering \includegraphics[width = \columnwidth]{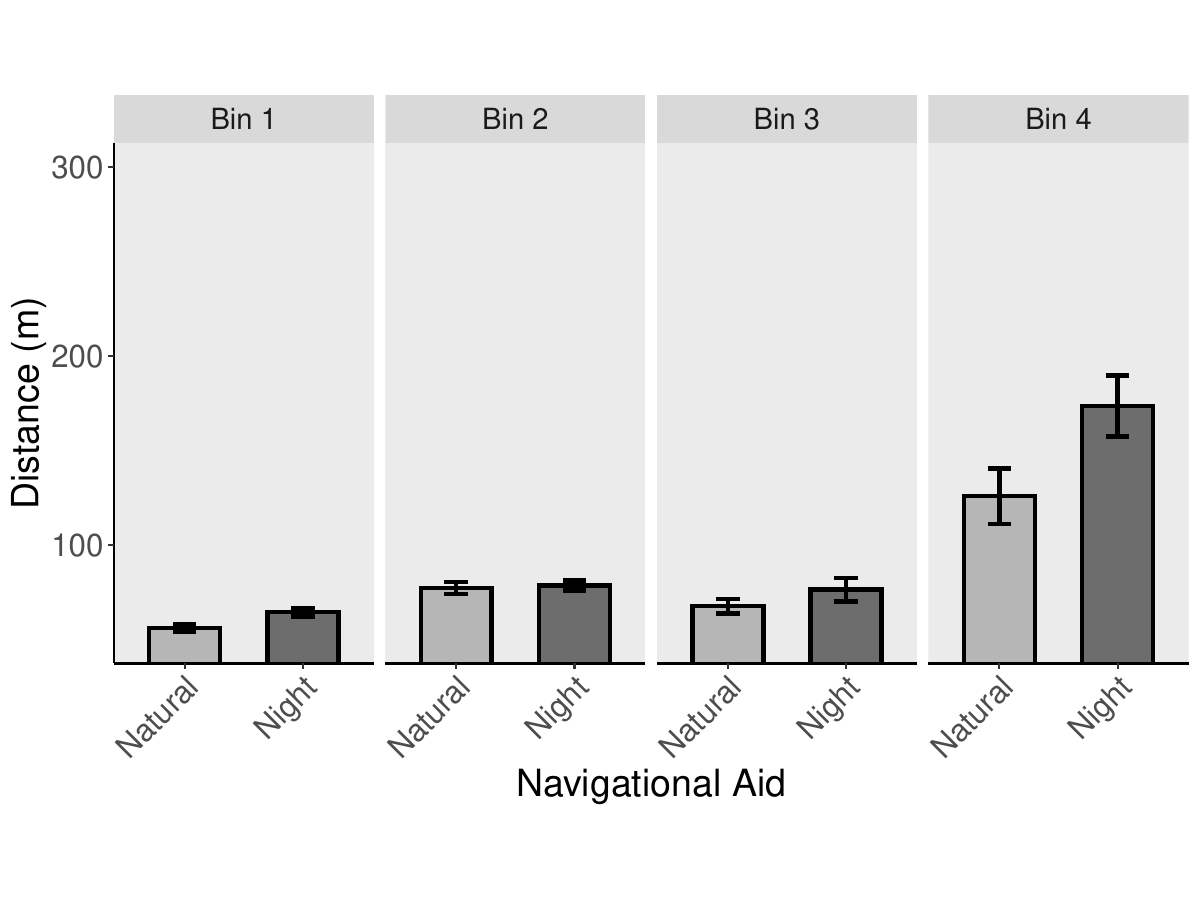}
\caption{Total distance traveled in meters plotted as a function of lighting condition and bin. Participants traveled a significantly longer distance at night compared to the natural light condition in the first and last bin.}
\Description{Total distance traveled in meters plotted as a function of lighting condition and bin. Participants traveled a significantly longer distance at night compared to the natural light condition in the first and last bin.}
\label{fig:dist_light}
\end{figure}

The ANOVA conducted on distance traveled revealed that $F$(1,22) = 8.32, $p$ < 0.01, $\eta_{}^{\mathrm{2}}$ = .27, $large$. There was no interaction between lighting and aid ($F$(3, 66) = 1.64, $p$ = 0.19, $\eta_{}^{\mathrm{2}}$ = .069). There was an interaction between lighting and bin (plotted in Figure \ref{fig:dist_light}), $F$(1.31,28.78) = 3.96, $p$ = 0.046, $\eta_{}^{\mathrm{2}}$ = .15, $large$, such that participants traveled a longer distance in the night compared to the natural lighting condition only in the first and the final bin; bin 1 [$t$(22) = -2.51, $p$ = 0.02, $d$ = -1.02, $large$]; bin 4 [$t$(22) = -2.42, $p$ = 0.024, $d$ = -0.98, $large$]. These results suggest that when participants were both adjusting to the task in the initial bin and searching for the final gems, they had greater difficulty finding gems in the night relative to the natural light condition. There was no three-way interaction $F$(2.94, 64.79) = 1.90, $p$ = 0.14, $\eta_{}^{\mathrm{2}}$ = .080. 

Two additional ANOVAs were used to examine whether lighting and aid conditions impacted both audio accuracy and mean response time. The ANOVA conducted on audio accuracy revealed no main effect of lighting condition $F$(1,22) = .26, $p$ = 0.61, $\eta_{}^{\mathrm{2}}$ = .012. There was also no interaction between lighting and aid $F$(3,66) = .52, $p$ = 0.67, $\eta_{}^{\mathrm{2}}$ = .023. The ANOVA on mean response time also revealed no main effect of lighting $F$(1,22) = .12, $p$ = 0.74, $\eta_{}^{\mathrm{2}}$ = .005. The interaction between lighting and aid was not significant $F$(2.06,45.32) = 0.80, $p$ = 0.46, $\eta_{}^{\mathrm{2}}$ = .035. 
}

% \blue{
% \section{Additional Results - Global Behavioral Metrics}
% \label{sec:appendix-behavioral}

% In addition to the analysis of head rotation and walking speed, we  performed an analysis of these two metrics accumulated over the duration of the bin/trial - accumulated head rotation and distance traveled, respectively. Although the results of these analyses mostly tracked with the time spent in each trial, we believe they are still informative and offer additional insights into participants' search behavior and therefore discuss them here. 
% }

\end{document}